\documentclass[12pt,draftcls,onecolumn]{IEEEtran}
\usepackage{mathrsfs}
\usepackage{amssymb}
\usepackage{cite}
\usepackage{graphicx}
\usepackage{color}

\def\rep#1{(\ref{#1})}

\newcommand{\R}{\mathbb{R}}

\def\send#1#2{\stackrel{#1}{\hbox to #2{\rightarrowfill}}}
\def\-{\!\!\!\!\!-}

 \def\qed{ \rule{.1in}{.1in}}
\def\eq#1{\begin{equation}#1\end{equation}}

\newcommand{\rank}{{\rm rank\;}}

\def\scr#1{{\cal #1}}

\newcommand{\matt}[1]{\left[ \matrix{#1} \right]}

\newcommand{\dfb}{\stackrel{\Delta}{=}}

\newtheorem{theorem}{Theorem}
\newtheorem{lemma}{Lemma}

\newtheorem{proposition}{Proposition}
\newtheorem{corollary}{Corollary}

\newcounter{seqn}[equation]
\def\theseqn{\arabic{equation}\alph{seqn}}

\def\endseqn{\eqno \@seqnnum
$$\ignorespaces}
\def\@seqnnum{(\theseqn)}

\newskip\mcentering \mcentering=0pt plus 1000pt minus 1000pt

\def\meqalignno#1{
\halign to\displaywidth{
    \hbox to 0pt{\kern\displaywidth\llap{$##$}\hss}\tabskip=\mcentering
    &\hfil$\displaystyle{##}$\tabskip=\mcentering
   &&$\displaystyle{{}##}$\hfil\tabskip=\mcentering
    \crcr
    #1\crcr}}

\def\rep#1{(\ref{#1})}

\def\eq#1{\begin{equation}#1\end{equation}}

\def\dspace{\multiply\normalbaselineskip 150
		  \divide\normalbaselineskip 100 \normalbaselines
		  \csname @@normalbaselineskip\endcsname\normalbaselineskip}
\def\sspace{\multiply\normalbaselineskip 200
		 \divide\normalbaselineskip 300 \normalbaselines
		 \csname @@normalbaselineskip\endcsname\normalbaselineskip}
\def\sdspace{\multiply\normalbaselineskip 160
		 \divide\normalbaselineskip 150 \normalbaselines
		 \csname @@normalbaselineskip\endcsname\normalbaselineskip}

\def\@{\tilde}

\def\3dot#1{\buildrel\textstyle...\over#1}

\begin{document}

\title{ A Distributed Algorithm  for Solving a Linear Algebraic
Equation\thanks{The authors  thank Daniel Spielman and  Stanley Eisenstat, Department of Computer Science, Yale University  for useful  discussions which have contributed to this work.
An abbreviated version of this paper was presented at the 51st Annual Allerton Conference on Communication, Control, and Computation
\cite{allerton2013}.
  This work was supported by the US
Air Force Office of Scientific Research and by the National Science Foundation. Shaoshuai Mou  is at MIT,  A. Stephen
 Morse is at Yale University and Ji Liu is at the University of Illinois, Urbana-Champaign. Emails: {\tt \small smou@mit.edu,  jiliu@illinois.edu,
as.morse@yale.edu. } Corresponding author: Shaoshuai Mou.} }
\author{Shaoshuai Mou \hspace{.5in} Ji Liu \hspace{.5in} A. Stephen Morse}

\markboth{ IEEE Transactions on Automatic Control, Accepted. }{Shell \MakeLowercase{\textit{et al.}}: Bare Demo of
IEEEtran.cls for Journals}

\maketitle
\begin{abstract} A distributed algorithm is described for solving a  linear algebraic equation of the form $Ax=b$
assuming the equation has at least one solution.  The equation is simultaneously
solved by $m$ agents
assuming each agent knows only a subset of the rows of the partitioned matrix $\matt{A &b}$,
 the current estimates  of the equation's solution generated by its neighbors, and nothing more. Each agent
recursively updates its estimate  by utilizing the current estimates
  generated by each of  its neighbors.
  Neighbor relations are characterized by a  time-dependent directed  graph $\mathbb{N}(t)$ whose
 vertices correspond to agents and whose arcs depict neighbor relations.
  It is   shown that for
  any  matrix $A$  for which the equation has a solution and any
  sequence of ``repeatedly jointly strongly connected graphs'' $\mathbb{N}(t)$, $t=1,2,\ldots$, the algorithm causes
 all agents' estimates to converge exponentially fast  to the same solution to $Ax=b$. It is also shown  that
 the neighbor graph sequence must actually be  repeatedly jointly strongly connected if exponential convergence is to be assured. A worst case convergence rate bound is derived for the case when $Ax=b$ has a unique solution.  It is demonstrated that with minor modification, the  algorithm can track the solution to $Ax = b$,
 even if $A$ and $b$ are changing with time, provided the rates of change of  $A$ and $b$  are sufficiently small.
   It is also shown that in the absence
 of communication delays, exponential
   convergence to a solution occurs  even if the times at which each agent updates its estimates are
   not synchronized with the update times of its neighbors. A modification of the algorithm is outlined which enables it to obtain a least squares solution to $Ax=b$ in a distributed manner, even if $Ax=b$ does not have a solution. \end{abstract}

\begin{IEEEkeywords}
Autonomous Systems; Distributed Algorithms; Linear Equations.
\end{IEEEkeywords}

\IEEEpeerreviewmaketitle

\section{Introduction }\newcommand{\rrr}[1]{\textcolor{red}{#1}}
Certainly the most well known and probably the most  important of all
numerical computations involving real numbers is solving a system of linear algebraic equations.
Efforts  to develop  distributed  algorithms to solve such systems have been under way for a long time
especially in the parallel processing community where the main objective is to achieve efficiency by
 somehow decomposing  a large system of linear equations into smaller ones  which can be solved on
  parallel processers more accurately or faster than direct solution of the original
  equations would allow \cite{margaris,koc,andersson,sor,kaczmarz}.
In some cases, notably in sensor networking \cite{xxx,infotheory} and some filtering applications \cite{khan},
the need for distributed processing  arises naturally
because processors onboard  sensors or robots are physically separated from each other.  In addition,
 there are typically
communication constraints which limit the flow of information across a robotic or sensor network
 and consequently preclude
centralized processing, even if efficiency is not the central issue. It is with these thoughts
 in mind that we are led to consider  the following  problem.

\section{ The Problem}

We are interested in a network of $m>1$ \{possibly mobile\} autonomous agents
 which are able to  receive information from  their ``neighbors'' where by a {\em neighbor} of agent
  $i$ is meant any other agent within agent $i$'s reception range. We write $\scr{N}_i(t)$ for the labels
  of agent $i$'s neighbors at time $t$,  and we always take agent $i$ to be a neighbor of itself.
  Neighbor relations at time $t$ can be conveniently
  characterized by a directed graph $\mathbb{N}(t)$  with $m$ vertices and a set of
   arcs defined so that
   there is an arc in $\mathbb{N}(t)$
   from vertex $j$ to vertex $i$ just
   in case agent $j$ is a neighbor of agent $i$ at time $t$. Thus  the  directions
    of arcs represent the directions of
   information flow.
Each agent $i$ has a real-time  dependent state vector $x_i(t)$ taking values in
$\R^n$,  and we assume that the information agent
 $i$ receives from neighbor $j$ at
  time $t$ is $x_j(t)$.
  We also assume that agent $i$
 knows a pair of real-valued  matrices $(A_i^{n_i\times n},b_i^{n_i\times 1})$.
The problem of interest is to devise local algorithms,
 one for each agent, which will enable all $m$ agents to iteratively and asynchronously compute
   solutions
 to the linear equation $Ax=b$
where $A = {\rm column}\;\{A_1,A_2,\ldots,A_m\}_{\bar{n}\times n} $, $b= {\rm column}
\;\{b_1,b_2,\ldots,b_m\}_{\bar{n}\times n}$
and  $\bar{n} =\sum_{i=1}^m n_i$. We shall require these solutions to be exact up to  numerical round off and communication  errors. In the first part of this  paper we will focus
  on the  synchronous case
 and we will assume
that  $Ax=b$  has
 a solution  although we will not require it to be unique. A restricted version of the
  asynchronous problem in which communication delays are ignored,  is addressed in \S\ref{asynch};
   a more general version  of the asynchronous problem in which communication delays are explicitly taken into account, is
 treated
 in  \cite{asylineareqn}.

The problem just formulated can  be viewed as a {\em distributed parameter estimation problem} in which
the $b_i$ are
{\em observations}
available to the sensors and $x$ is a parameter to be estimated. In this setting, the observation equations
 are sometimes of the form $b_i = A_ix +\eta_i$ where $\eta_i$ is a
 term modeling measurement noise \cite{infotheory}.
The most widely studied  version of the problem is  when $m=n$, the $A_i$ are linearly
independent row vectors $a_i$, the $ b_i$ are  scalars,
and $\mathbb{N}(t)$ is a
constant, symmetric  and strongly connected graph.
For this version of the problem, A is  therefore an  $n\times n$  nonsingular matrix,
$b$ is an $n$ vector and agent $i$ knows the state $x_j(t)$ of each of its neighbors
 as well as its own state. The problem in this case is thus for each agent $i$
 to compute $A^{-1}b$, given $a_i, b_i$ and $x_j(t),\;j\in\scr{N}_i,\;\;t \geq 0$.
 In this form, there are several classical parallel algorithms which address closely related problems. Among
  these are Jacobi iterations \cite{margaris}, so-called ``successive over-relaxations''
   \cite{sor} and the classical Kaczmart method \cite{kaczmarz}.
   Although these are parallel algorithms, all
rely on ``relaxation factors'' which cannot be determined  in a distributed way unless
one makes special  assumptions about $A$. Additionally, the implicitly defined neighbor graphs for
these algorithms are generally strongly  complete; i.e., all processors can communicate with each other.

This paper breaks new ground by providing an algorithm which is
 \begin{enumerate}
 \item\label{c1} applicable to {\em any} pair of real  matrices $(A,b)$  for which $Ax=b$ has at least one solution.
 \item\label{c2} capable of finding a solution at least exponentially fast \{Theorem \ref{mainer}\}.
 \item\label{c3} applicable to the {\em largest}  possible class of  time-varying  directed neighbor   graphs $\mathbb{N}(t)$ for which exponential convergence can be assured \{Theorem \ref{mainer.nes}\}.
\item\label{c4}  capable of finding a solution to $Ax = b$ which, in the absence of  round off and communication errors, is  exact.
 \item\label{c5} capable of finding a solution using at most an $n$ dimensional state vector received at each clock time from each of its neighbors.

\item\label{c6} applicable without imposing restrictive or unrealistic requirements such as (a) the assumption that each agent is  constantly  aware of an upper bound on the number of neighbors of each of its neighbors or (b) the assumption that all agents are able to share the same time-varying step size.

    \item\label{c7}    capable of operating asynchronously.

\end{enumerate}


An obvious approach to the problem we've posed
 is to reformulate  it as a distributed optimization problem and then try to use existing algorithms such as those in \cite{nedic,KM12Allerton,DJJ12CDC,DJJ14TAC,nedic2,KM12CDC,JAM12TAC,TAA14TAC,AA13CDC,cortes,AA12Allerton}  to obtain a solution. Despite  the fact that there is a large literature on distributed optimization, we are not aware of any distributed optimization algorithm which, if applied to the problem at hand, would
 possess all of the attributes mentioned above, even if the capability of functioning asynchronously were not on the list. For the purpose of solving the problem of interest here, existing algorithms are deficient in various ways. Some can only find approximate  solutions with bounded errors \cite{nedic}; some are only applicable to networks with bi-directional communications
  \{ie, undirected graphs\} and/or networks with fixed graph topologies \cite{KM12Allerton,DJJ12CDC,DJJ14TAC,JAM12TAC}; many require all agents to share a common, time varying step size \cite{nedic2,KM12CDC,KM12Allerton,JAM12TAC,DJJ14TAC,TAA14TAC,AA13CDC}; many introduce an additional scalar or vector state \cite{AA13CDC,cortes, DJJ12CDC, DJJ14TAC,TAA14TAC,KM12CDC,AA12Allerton} for each agent to update and transmit; none have been shown  to generate solutions which converge  exponentially fast, although it is plausible that some may
  exhibit exponential convergence when applied to the type of quadratic optimization problem one would
  set up  to solve  the  linear equation which is of interest here.


   One limitation common to many distributed optimization algorithms is the requirement that each agent must be aware of an upper bound on the number of neighbors of each of its neighbors. This means that there must be bi-directional communications between agents. This requirement can be quite restrictive, especially if neighbor relations change with time. The requirement stems from the fact that
    most distributed optimization algorithms  depend on  some form of ``distributed averaging.''
     {\em Distributed averaging} is a special type of  consensus seeking for which the goal is for all $n$ agents to ultimately compute the  average  of the initial values of their consensus variables. In  contrast, the goal of  consensus seeking is for all agents to ultimately agree on a common value of their consensus variable, but that  value need not be  the average of their initial values. Because distributed
   averaging is a special form of consensus seeking, the methods used to obtain a distributed average are  more specialized than those needed to  reach a consensus.
 There are  three different approaches to distributed averaging: linear iterations\cite{boydweight,xxx}, gossiping\cite{boyd05,deterministic}, and
   double linear iterations \cite{doublelinear} which are also known as push-sum algorithms  \cite{kempe,doub1,KM12CDC} and scaled agreement algorithms \cite{olshevsky.siam}.

    Linear iterations for distributed averaging can be modeled as a linear recursion equation in which the  \{possibly time-varying\} update matrix must be doubly stochastic \cite{boyd05}.
    The doubly stochastic matrix requirement cannot be satisfied without assuming that each agent knows  an upper bound on the number of neighbors of each of its neighbors. A recent exception to this is the paper \cite{AC14TAC} where the idea is to learn weights within the requisite doubly stochastic matrix in an asymptotic fashion. Although this idea is interesting, it also adds complexity to the  distributed averaging  process; in addition, its applicability is limited to time invariant graphs.

     Gossiping is a very widely studied approach to distributed averaging in which each agent is  allowed to average its consensus variable with at most one other agent at each clock time. Gossiping protocols can lead to deadlock unless specific precautions are taken to insure that they do not and these precautions generally lead to fairly complex  algorithms \cite{deterministic} unless one is willing to accept probabilistic solutions.

       Push-sum algorithms  are based on a quite clever idea first apparently proposed by  in \cite{kempe}.   Such algorithms are somewhat more complicated than linear iterations, and generally require more data to be communicated between agents. They are however attractive because, at least for some implementations, the requirement that each agent know the number of neighbors of each of its neighbors is avoided  \cite{doublelinear}.

Another approach to the problem we have posed is to reformulate it as a least squares problem.   Distributed algorithms capable of solving the least squares problem have the advantage of being applicable to $Ax=b$ even when this equation has no solution. The authors of \cite{lu1,lu3} develop several algorithms for solving this type of problem and give sufficient
  conditions for them to work correctly; a limitation of their algorithms is that each agent is assumed to know the coefficient matrix $A_j$ of each of its neighbors.  In \cite{tron},  it is noted that the distributed least squares problem can be solved by using distributed averaging to compute the average of the matrix pairs
  $(A_i'A_i, A_i'b_i)$.  The downside of this very clever idea  is that the amount of data to be communicated between agents does not scale well as the number of agents increases.  In \S\ref{LS} of this paper an alternative approach to the distributed  least squares problem  is briefly outlined; it too has scaling problems,
  but also  appears to have the potential of admitting a modification which will to some extent overcome the scaling problem.

 Yet another  approach to the problem of interest in this paper, is to view it as a consensus
problem in which the goal is for all $m$ agents to ultimately attain the same value for their states subject
 to the
requirement that each $x_i$ satisfies the convex constraint $A_ix_i = b_i.$
An algorithm for solving a large class of constrained consensus
problems of this type in a synchronous manner, appears in \cite{nedic2}.
Specialization of that algorithm to the problem of interest here,
yields an algorithm similar to synchronous version of the algorithm
which we will consider. The principle difference between the two -
apart from correctness proofs  and claims about speed of convergence - is that the algorithm stemming from
\cite{nedic2} is based on distributed averaging and  consequently relies on convergence properties of doubly stochastic matrices
whereas the synchronous version of the algorithm developed in
this paper does not. As a consequence, the algorithm stemming from \cite{nedic2} cannot be
implemented without assuming that each agent knows as a function of time, at least an
upper bound on the number of neighbors of each of its current
neighbors, whereas the algorithm under consideration here can. Moreover, limiting the consensus updates to distributed averaging  via  linear iterations  almost certainly
limits the possible convergence rates which might be achieved, were one not constrained by the special structure of doubly stochastic matrices.  We see no reason  at all to limit the algorithm we are discussing to doubly stochastic matrices  since, as this paper demonstrates, it is not necessary to. In addition, we mention
 that a convergence proof
 for the constrained consensus algorithm proposed in \cite{nedic2} which avoids doubly stochastic matrices   is claimed to have been developed in \cite{linwen}  but the correctness of the proof presented there is far from clear.

 Perhaps the most important difference between the results of \cite{nedic2} and the results to be presented here  concerns speed of convergence. In this paper  exponential convergence is established  for any sequence of repeatedly strongly connected neighbor graphs.  In \cite{nedic2}, asymptotic convergence is proved under the same neighbor graph conditions, but {\em exponential} convergence  is only proved in the special case when the neighbor graph  is  fixed and complete. It is not  obvious how to modify the analysis given in \cite{nedic2} to obtain a proof of  exponential convergence under more relaxed conditions.

In contrast with earlier work  on  distributed optimization and distributed consensus, the  approach taken in this paper is based on a simple observation,
inspired by \cite{nedic2}, which has the potential on being applicable to a broader class of problems than being considered here.
Suppose that one is interested in devising a distributed algorithm which can cause all members of a group of
$m$ agents  to find a solution $x$ to the system of equations $\alpha_i (x) = 0,\;i\in\{1,2,\ldots,m\}$ where
$\alpha_i:\R^n\rightarrow \R^{n_i}$  is a ``private'' function know only to agent $i$. Suppose each agent $i$
is able to find a solution $x_i$ to its private equation $\alpha_i(x_i) = 0$, and in addition, all of the $x_i$ are the same. Then all $x_i$ must satisfy $\alpha_j (x_i) = 0,\;j\in\{1,2,\ldots,m\}$ and thus each constitutes a solution to the problem. Therefore to solve such a problem, one should try to craft an algorithm which, on the one hand causes
each  agent  to  satisfy its own private equation and on the other causes all agents to reach a consensus.
We call this   the {\em agreement principle}. We don't know if it has been articulated   before although it has been used  before without special mention  \cite{lineareqnf}. As we shall see, the agreement principle is the basis
for three different versions of the problem we are considering.

\section{The Algorithm}\label{alg}

 Rather than go through the intermediate step of reformulating
 the
 problem under consideration   as an optimization problem or as a constrained consensus problem, we shall approach the problem
directly in accordance with the agreement principle.   This was already   done in  \cite{lineareqnf} for the case  when neighbors do not change and the algorithm obtained was the same one as the one we are about to develop here.
 Here  is  the idea  assuming that all agents  act synchronously. Suppose time is discrete in that $t$ takes values in $\{1,2,\ldots\}$.
Suppose that at each time $t\geq 1$, agent $i$ picks as a preliminary estimate of a solution to $Ax=b$, a solution
 $z_i(t)$ to $A_ix=b_i$.
  Suppose that
$K_i$ is a basis matrix for the kernel of $A_i$.
If we set $x_i(1) = z_i(1)$ and
restrict the updating  of $x_i(t)$ to iterations of the form
$x_i(t+1) = z_i(t) + K_iu_i(t),\;\;t\geq 1$,
then no matter what $u_i(t)$ is,  each $x_i(t)$ will  obviously satisfy $A_ix_i(t) = b_i,\;t\geq 1$.
Thus, in accordance with the agreement principle,  all we need to do  to solve the problem is to come up with
 a good way to choose the $u_i$ so that a consensus is ultimately reached. Capitalizing  on what is known
 about consensus algorithms
   \cite{reachingp1,vicsekmodel,blondell},
  one would like to choose $u_i(t)$ so that
  $x_i(t+1) = \frac{1}{m_i(t)}\left(\sum_{j\in\scr{N}_i(t)}x_j(t)\right )$
  where $m_i(t)$ is the number of neighbors of agent $i$ at time $t$,
but this is impossible to do  because  $-z_i(t)+\frac{1}{m_i(t)}\sum_{j\in\scr{N}_i(t)}x_j(t)$ is not typically
in the image of $K_i$. So instead one might try
choosing each $u_i(t)$ to minimize the difference
 $\left (z_i(t)+K_iu_i(t)\right ) - \frac{1}{m_i(t)}\left(\sum_{j\in\scr{N}_i(t)}x_j(t)\right )$ in the least squares sense.
Thus   the idea  is to choose $x_i(t+1)$
to satisfy $A_ix_i(t+1) = b_i$  while at the same time making  $x_i(t+1)$  approximately
 equal to the average of agent $i$'s neighbors'  current estimates of the solution to $Ax=b$.
Doing this leads at once to an iteration for
 agent $i$
  of the form

\eq{x_i(t+1) = z_i(t)- \frac{1}{m_i(t)}P_i\left (  m_i(t)z_i(t)-\sum_{j\in\scr{N}_i(t)}x_j(t)\right ),\;\;t\geq 1\label{a1nn}}

\noindent where $P_i$ is the readily computable  orthogonal projection on the kernel of  $A_i$.
Note right away that the algorithm  does not involve   a relaxation factor and is totally distributed.
While the intuition upon which this algorithm is based is  clear, the algorithm's correctness
 is not.

  It is easy to see that $(I-P_i)z_i(t)$ is fixed no matter what  $z_i(t)$ is, just  so long as it is a solution to $A_ix = b_i$ .  Since $x_i(t)$ is such a solution, \rep{a1nn}  can also be written as
\eq{x_i(t+1) = x_i(t)- \frac{1}{m_i(t)}P_i\left (  m_i(t)x_i(t)-\sum_{j\in\scr{N}_i(t)}x_j(t)\right ),\;\;t\geq 1\label{a1}}
 and it is  this form which we shall study. Later in \S\ref{tracking} when we focus on a generalization of the problem in which $A$ and $b$ change  slowly with time,  the corresponding generalizations of \rep{a1nn} and \rep{a1}
 are not quite equivalent and it will  be more convenient to focus on the generalization corresponding to \rep{a1nn}.

As mentioned in the preceding section, by specializing  the  constrained consensus problem treated in \cite{nedic2}
to the problem
 of interest here, one can obtain  an update rule similar to  \rep{a1}. Thus the arguments in \cite{nedic2}
can  be used to
 establish asymptotic convergence in the case of  synchronous operation.  Of course using the powerful but lengthy and
 intricate proofs developed in \cite{nedic2}  to address  the  specific constrained consensus problem posed here, would seem to be a
 round about way of analyzing  the problem, were there available  a direct and more transparent  method.
One of the  main contributions of this paper is to provide just such a method.  The method
   closely parallels the well-known  approach to unconstrained consensus problems based on nonhomogeneous Markov chains \cite{seneta,vicsekmodel}.
The standard unconstrained consensus problem is typically studied by looking at the   convergence properties
of infinite products of $S_{m\times m}$ stochastic matrices. On the other hand,    the
problem posed in this paper is studied by looking at  infinite products of  matrices of the form
$P(S\otimes I)P$ where $P$ is a block diagonal matrix of $m$, $n\times n$ orthogonal matrices, $S$ is an $m\times m$ stochastic  matrix,
 $I$ is the $n\times n$ identity, and $\otimes$ is the Kronecker product. For the standard unconstrained   consensus problem,
the relevant measure of the distance of a stochastic matrix $S$ from the desired limit of a rank  one stochastic matrix  is the
infinity matrix semi-norm \cite{deterministic} which is also the same thing as the well known coefficient of
ergodicity \cite{seneta}. For the problem posed in this paper, the relevant measure of the distance of  a matrix of the form
$P(S\otimes I)P$
 from the desired limit of the zero matrix, is a somewhat unusual but especially useful concept called a  ``mixed-matrix'' norm \S \ref{U}.

\section{Organization} The remainder of this paper is organized as follows.
 The discrete-time  synchronous case is treated first. We begin in Section \ref{mrs} by stating conditions
  on the sequence of neighbor graphs $\mathbb{N}(1), \mathbb{N}(2),\ldots $ encountered along a ``trajectory,'' for the overall distributed algorithm
 based on \rep{a1} to converge exponentially fast to a solution to $Ax=b$. The conditions on the neighbor graph sequence are both sufficient \{Theorem \ref{mainer}\} and necessary \{Theorem \ref{mainer.nes}\}. A worst case
 geometric convergence rate is then given \{Corollary \ref{4th}\} for the case when $Ax=b$ has a unique solution.


  Analysis of the synchronous case is carried out in \S\ref{A}. After developing the relevant linear iteration \rep{pup}, attention is focused in \S\ref{U} on proving that  repeatedly jointly strongly connected neighbor graph sequences are sufficient for exponential convergence.  For the case when $Ax=b$ has a unique solution, the problem reduces to
 finding conditions \{Theorem \ref{nmeets}\} on an infinite sequence of $m\times m$ stochastic matrices $S_1,S_2,\ldots $ with positive diagonals
 under which
 an infinite sequence of matrix products of the form  $(P(S_k\otimes I)P)(P(S_{k-1}\otimes I)P)\cdots (P(S_1\otimes I)P),\;k\geq 1$ converges to the zero matrix.
 The problem is similar to problem
 of determining conditions on an infinite sequence  of $m\times m$ stochastic matrices $S_1,S_2,\ldots $ with positive diagonals
 under which
 an infinite sequence of matrix products of the form  $(S_kS_{k-1}\cdots S_1),\;k\geq 1$ converges to
 a rank one stochastic matrix.
The latter problem is addressed in the standard consensus literature
by exploiting several facts:

 \begin{enumerate}
 \item The induced infinity matrix semi-norm \cite{deterministic} \{i.e., the coefficient of ergodicity \cite{seneta}\} is sub-multiplicative
on the set of $m\times m$ stochastic matrices.
 \item  Every  finite product of stochastic matrices
is non-expansive in the induced infinity matrix semi-norm \cite{deterministic}.
  \item \label{er1} Every sufficiently long product of stochastic matrices with positive diagonals is a semi-contraction in the infinity semi-norm provided the graphs of the stochastic matrices appearing in the product are all rooted\footnote{A directed graph is {\em rooted} if it contains  at least one vertex  $r$ from which, for each vertex $v$ in the graph, there is a directed path from $r$ to $v$ .} \cite{reachingp1,reachingp2, deterministic}.
      \end{enumerate}
      \noindent There are parallel results for the problem of interest here: \begin{enumerate}
      \item The mixed matrix norm defined by \rep{mmn} is sub-multiplicative on $\R^{mn\times mn}$  \{Lemma \ref{sub}\}. \item  Every finite matrix product of the form
$(P(S_k\otimes I)P)(P(S_{k-1}\otimes I)P)\cdots (P(S_q\otimes I)P)$  is non-expansive in the mixed matrix norm \{Proposition \ref{ve}\}. \item \label{er2} Every  sufficiently product of such matrices is a contraction in
the mixed matrix norm provided  the stochastic matrices appearing in the product have positive diagonals and graphs which are all strongly connected \{Proposition \ref{ppo}\}.

  \end{enumerate}

 While there are many similarities between the consensus problem and the problem under consideration here,  one important difference is that the
 set of $m\times m$ stochastic matrices is closed under multiplication whereas the set of matrices of the form  $(P(S\otimes I)P)$ is not. To deal with this, it is necessary to introduce the idea of a
 ``projection block matrix'' \S\ref{PBM}. A projection block matrix is a partitioned matrix whose
 specially structured blocks are called
 ``projection matrix polynomials''  \S\ref{PRO}. What is important about this concept is that the
   set of projection block matrices is closed under multiplication  and  contains every matrix product of the form $(P(S_k\otimes I)P)(P(S_{k-1}\otimes I)P)\cdots (P(S_q\otimes I)P)$.  Moreover, it is possible to give conditions under which a projection block matrix is a contraction in the mixed matrix norm \{Proposition \ref{ve}\}. Specialization of this result yields a characterization of the class of matrices  of the form $(P(S_k\otimes I)P)(P(S_{k-1}\otimes I)P)\cdots (P(S_q\otimes I)P)$ which are contractions \{ Proposition \ref{ppo}\}.  This, in turn is used to prove Theorem \ref{nmeets} which is the main technical result of the paper.

  The proof of Theorem \ref{mainer} is carried out in two steps.  The case
  when $Ax=b$ has a unique solution is treated first. Convergence in this case is an immediate consequence of  Theorem \ref{nmeets}. The general  case   without the assumption of uniqueness is treated next.  In this case, Lemma \ref{vlad}  is used to decompose the problem into two parts - one to which the  results for the uniqueness case are directly applicable and the other  to which standard unconstrained consensus results are  applicable.

  It is well known that the necessary condition for a standard  unconstrained consensus algorithm to generate an exponentially convergent solution is that the sequence of neighbor graphs encountered be ``repeatedly jointly rooted'' \cite{luc}.
  Since a ``repeatedly jointly strongly connected sequence'' is always a repeatedly jointly rooted sequence, but not conversely, it may at first glance seem surprising that for the problem under consideration in this paper,
   repeatedly jointly strongly connected sequences are in fact necessary for exponential convergence. Nonetheless
   they are and a proof of this claim is given in Section \ref{nec}. The proof relies on the concept of an ``essential
   vertex''   as well as the idea of ``a mutual reachable equivalence class.'' These ideas can be found in
   \cite{seneta} and \cite{gallager} under different names.

 Theorem \ref{nmeets} is proved in \S \ref{gal}. The proof relies heavily on  a number of concepts mentioned earlier
     including  the mixed matrix norm,  projection matrix polynomials \{\S \ref{PRO}\}, and projection block matrices \{\S\ref{PBM}\}. These concepts also  play an important role in \S\ref{CR}  where
the  worst case convergence rate stated in Corollary  \ref{4th} is justified.
 To underscore  the importance of exponential convergence,  it is explained  in \S\ref{tracking} why
   that with minor modification, the  algorithm we have been considering can track the solution to $Ax = b$,
  if $A$ and $b$ are changing with time, provided the rates of change of  $A$ and $b$  are sufficiently small.
 Finally, the
    asynchronous version of the problem  is  addressed in Section \ref{asynch}.

A limitation of the algorithm we have been discussing is that it is only applicable to linear equations for which there are solutions. In  \S \ref{LS} we explain how to modify the algorithm so that it can obtain least squares solutions to $Ax = b$ even in the case when $Ax = b$ does not have a  solution.  As before, we  approach the problem using standard consensus concepts rather than the more restrictive  concepts  based on distributed averaging.

\subsection{Notation}  If $M$ is a matrix,  $\scr{M}$ denotes its column span.  If $n$ is a positive integer,
$\mathbf{n} = \{1,2,\ldots, n\}$. Throughout this paper
$\scr{G}_{sa}$ denotes the set of all directed graphs with
  $m$ vertices 
    which have self-arcs at all vertices.
The graph of an $m\times m$ matrix  $M$ with nonnegative entries is an $m$ vertex directed
 graph $\gamma(M)$  defined so that $(i,j)$ is an arc from $i$ to $j$  in the graph just in case
  the $ji$th entry of
 $M$ is nonzero. Such a graph is in $\scr{G}_{sa}$ if and only if all diagonal entries of
  $M$ are positive.

\section{Synchronous Operation}\label{mrs}

Obviously conditions for convergence of the $m$ iterations defined by \rep{a1} must depend on   neighbor graph connectivity. To make precise just what is meant by connectivity in the present context, we need
 the idea of ``graph composition''    \cite{reachingp1}.
By the
 the {\em composition} of a directed graph $\mathbb{G}_p\in\scr{G}_{sa}$
with a directed graph $\mathbb{G}_q\in\scr{G}_{sa}$,  written $\mathbb{G}_q\circ\mathbb{G}_p$
is meant that directed graph in $\scr{G}_{sa}$ with
 arc set defined  so that $(i, j)$ is an arc in the
composition just in case there is a vertex $k$ such that $(i, k)$
 is an arc in $\mathbb{G}_p$ and $(k, j)$
is an arc in $\mathbb{G}_q$ .  It is clear that $\scr{G}_{sa}$   is closed under composition and
composition is an associative binary operation; because of this, the definition extends
unambiguously to any finite sequence of directed graphs   in $\scr{G}_{sa}$.
Composition is defined so that
for any pair of  nonnegative $m\times m$ matrices $M_1,M_2$, with graphs
 $\gamma(M_1),\gamma(M_2)\in\scr{G}_{sa}$,
 $\gamma(M_2M_1)$ =$\gamma(M_2)\circ\gamma(M_1)$.

To proceed,
let us agree to say that   an infinite  sequence of graphs $\mathbb{G}_1, \mathbb{G}_2,\ldots $  in $\scr{G}_{sa}$
is
{\em repeatedly   jointly strongly connected}, if  for some finite positive integers  $l$ and $\tau_0$ and
 each integer $k>0$, the composed graph
 $\mathbb{H}_k = \mathbb{G}_{kl +\tau_0-1}\circ\mathbb{G}_{kl+\tau_0-2}\circ  \cdots$ $\circ\mathbb{G}_{(k-1)l+\tau_0},$
is strongly connected. Thus if $\mathbb{N}_1, \mathbb{N}_2,\ldots $ is a sequence of neighbor graphs which is
repeatedly jointly strongly connected, then over each successive
 interval  of $l$  consecutive iterations starting at $\tau_0$, each proper subset of agents receives some  information
  from the rest. The first of the two  main results of this paper for synchronous operation is as follows.

\begin{theorem} Suppose each agent $i$ updates its state $x_i(t)$ according to  rule \rep{a1}.
If the sequence of neighbor graphs
 $\mathbb{N}(t),\; t\geq 1$, is repeatedly jointly strongly  connected, then
 there exists a positive constant
$\lambda<1$  for which
 all $x_i(t)$ converges to the same solution to $Ax=b$ as $t\rightarrow\infty$, as fast as $\lambda^t$ converges to $0$.
\label{mainer}\end{theorem}\vspace{.1in}

\noindent In the next section we explain why this theorem is true.

The idea of  a repeatedly jointly strongly connected sequence of graphs is the direct analog of the idea of a ``repeatedly jointly rooted'' sequence of graphs;  the repeatedly jointly rooted  condition, which is
weaker than the repeatedly jointly strongly connected condition,  is known to be not only a sufficient condition but also a necessary one on
an infinite sequence of neighbor graphs in $\scr{G}_{sa}$  for all agents in an unconstrained  consensus process to reach a consensus exponentially fast \cite{luc}.
The question then, is  repeatedly jointly strongly connected strong connectivity  necessary for exponential convergence of the $x_i$ to a solution to $Ax=b$? Obviously such a condition cannot be necessary in the special case when $A=0$ and \{and consequently $b=0$\}
because in the case the problem reduces to an unconstrained consensus problem. The repeatedly jointly strongly connected condition
also cannot be necessary
 if  a  distributed solution  to $Ax=b$  can be obtained
by only  a  proper subset of the full set of $m$ agents.
 Prompted by this, let  us agree to say
that agents with  labels in
$\scr{V} =\{i_1,i_2,\ldots, i_q\}\subset \mathbf{m}$  are {\em redundant} if any solution to the equations
$A_ix = b_i$ for  all $i$ in the complement of $\scr{V}$, is a solution to $Ax = b$. To derive an algebraic condition
 for redundancy, suppose that $z$ is a solution to $Ax=b$. Write $\bar{\scr{V}}$
 for the complement of $\scr{V}$ in $\mathbf{m}$.   Then any solution $w$ to the equations $A_ix=b_i,i\in\bar{\scr{V}}$
 must satisfy $w-z\in\bigcap_{i\in\bar{\scr{V}}}\scr{P}_i$,  where for $i\in\mathbf{m}$,  $\scr{P}_i = {\rm image}\;P_i$. Thus agents with labels in  $\scr{V}$ will be redundant
  just in case  $w-z\in\bigcap_{i\in\scr{V}}\scr{P}_i$.
Therefore agents with labels in  $\scr{V}$ will be redundant if and only if
$$\bigcap_{i\in\bar{\scr{V}}}\scr{P}_i\subset \bigcap_{i\in\scr{V}}\scr{P}_i.$$
We say that $\{P_1,P_2,\ldots, P_m\} $ is a {\em non-redundant set} if no such proper subset exists.  We can now state the second main  result of this paper for synchronous operation.

\begin{theorem} Suppose each agent $i$ updates its state $x_i(t)$ according to  rule \rep{a1}. Suppose in addition that $A\neq 0  $ and that
$\{P_1,P_2,\ldots,P_m\}$ is a non-redundant set. If there
 exists a positive constant
$\lambda<1$  for which
 all $x_i(t)$ converges to the same solution to $Ax=b$ as $t\rightarrow\infty$ as fast as $\lambda^t$ converges to $0$, then the sequence of neighbor graphs
 $\mathbb{N}(t),\; t\geq 1$, is repeatedly jointly strongly  connected.
\label{mainer.nes}\end{theorem}\vspace{.1in}

\noindent In the \S\ref{nec} we explain why this theorem is true.

For the case when $Ax=b$ has a unique solution and each of the neighbor graphs $\mathbb{N}(t),\;t\geq 1$ is strongly connected,   it is possible  to derive an explicit worst case
bound on the rate at which the $x_i$ converge. 
As will be explained at the beginning of \S \ref{U}, the uniqueness assumption is equivalent to the assumption that $\bigcap_{i\in\mathbf{m}}\scr{P}_i = 0$.  This and Lemma \ref{crack} imply that
the induced two-norm $|\cdot|_2$ of any finite product of the form $P_{j_1}P_{j_2}\cdots P_{j_k}$ is less than $1$, provided each of the $P_i,\;i\in\mathbf{m}$, occur in the product at least once. Thus if  $q\dfb (m-1)^2 $ and   $\scr{C}$  is the set of all such products of
length $q+1$, then $\scr{C}$ is compact and
\eq{\rho =\max_{\scr{C}}|P_{j_1}P_{j_2}\cdots P_{j_{q+1}}|_2\label{rate2}} is a number less than $1$.
So therefore is
\eq{\lambda =  \left (1-\frac{(m-1)(1-\rho)}{m^q}\right )^{\frac{1}{q}}.\label{rate}}
We are led to the following result.

\begin{corollary} Suppose that $Ax = b$ has a unique solution $x^*$. Let $\lambda$ be given by \rep{rate}.
 If each of the neighbor graphs $\mathbb{N}(t),\;t\geq 1$
mentioned in the statement of Theorem \ref{mainer} is strongly connected, then  all $x_i(t)$  converge to $x^*$ as $t\rightarrow 0$ as fast as $\lambda^t$ converges to $0$.
\label{4th}\end{corollary}\vspace{.1in}

\noindent A proof of this corollary will be given in  section \ref{CR}. The extension of this result to the case when $Ax=b$ has more than one solution can also be worked out, but this will not be done here. It is likely that $\rho$ can be related to a conditioning number for $A$, but this will not be done here.

In the consensus literature \cite{blondell},  researchers have also looked at
algorithms using convex combination rules rather than
 straight average rule which we have exploited here.
Applying such rules to the problem at hand
 leads to update equations of the more general  form
\eq{
x_i(t+1) = x_i(t)- P_i\left (  x_i(t)-\sum_{j\in\scr{N}_i(t)}w_{ij}(t)x_j(t)\right )\;\;\;\;i\in{\bf m}\label{soppp}}
where   the $w_{ij}(t)$ are nonnegative numbers  summing to $1$ and uniformly bounded from below by a positive constant.  The extension of the analysis  which follow
to encompass this generalization is straightforward. It should be pointed out however, that innocent  looking generalizations of these update laws which one might want to  consider, can lead to problems.  For example, problems can arise if the same value of $w_{ij}$ is not used to weigh all of the components of agent $j$'s  state in agent $i$'s update equation. To illustrate this, consider a network with a
  fixed two agent strongly connected  graph and $A_1 = \matt{1 & 1}$ and $A_2= \matt{-a & -1}$. Suppose agent $1$ uses weights $w_{1j} = .5.$ to weigh both components of $x_j, \;j\in\mathbf{2}$ but agent $2$
  weights the first components of state vectors $x_1$ and $x_2$ with $.25$ and $.75$ respectively while weighing
   the second components of both with $.5$.   A simple computation reveals that the spectral radius of the relevant update matrix for the state of the system determined by \rep{soppp} will  exceed $1$ for values of $a$ in the open interval $(.5, 1)$.


\section{Analysis}\label{A} In this section we explain why Theorems \ref{mainer} and \ref{mainer.nes}  are true. As a first step, we   translate the state  $x_i$  of \rep{a1} to a new shifted state $e_i$ which can be interpreted as is the error between $x_i$ and a solution to $Ax=b$;  as we shall see,  this simplifies the analysis.
Towards this end,
let $x^*$ be any solution to $Ax=b$.  Then  $x^*$ must satisfy $A_ix^* = b_i$ for  $i\in\mathbf{m}$.
 Thus if we define
 \eq{e_i(t) = x_i(t)-x^*,\;\;\;i\in\mathbf{m},\;\;t\geq 1\label{redefine}}
then $e_i(t)\in \scr{P}_i,\;t\geq 1$, because $\scr{P}_i = \ker A_i$. Therefore
$P_ie_i(t) = e_i(t),\;i\in\mathbf{m},\;\;t\geq 1$.
Moreover from \rep{a1},
$$e_i(t+1) = P^2_ie_i(t)- \frac{1}{m_i(t)}P_i\left
 (  m_i(t)P_ie_i(t)-\sum_{j\in\scr{N}_i(t)}P_je_j(t)\right )$$
 for $t\geq 1,\;\;i\in\mathbf{m}$,
 which simplifies to
\eq{e_i(t+1) = \frac{1}{m_i(t)}P_i\sum_{j\in\scr{N}_i(t)}P_je_j(t),\;\;t\geq 1,\;\;i\in\mathbf{m}.\label{na1}}
As a second step, we
 combine these $m$ update equations into one linear
recursion equation with state vector
 $e(t) = {\rm column}\{e_1(t),e_2(t),\ldots, e_m(t)\}$.  To accomplish this,  write
 $A_{\mathbb{N}(t)}$ for the adjacency matrix of $\mathbb{N}(t)$, $D_{\mathbb{N}(t)}$
 for the diagonal matrix whose $i$th diagonal entry is $m_i(t)$ \{$m_i(t)$ is  also the in-degree
 of vertex $i$ in $\mathbb{N}(t)$\}, and
 let  $F(t) =D^{-1}_{\mathbb{N}(t)}A'_{\mathbb{N}(t)}$.  Note that $F(t)$
  is  a stochastic matrix; in the literature it is sometimes  referred to as a {\em flocking matrix}.
It is straightforward to verify  that
\eq{e(t+1) =P(F(t)\otimes I)Pe(t),\;\;t\geq 1\label{pup}}
where
 $P$ is the $mn\times mn$ matrix $P = {\rm diagonal}\{P_1,P_2,\ldots, P_m\}$
and $F(t)\otimes I$ is the  $mn\times mn$   matrix which results  when each entry $f_{ij}(t)$ of $F(t)$ is replaced by
$f_{ij}(t)
$ times the
 $n\times n$ identity. Note that $P^2=P$ because each $P_i$ is idempotent.  We will use this fact without special mention in the sequel.

\subsection{Repeatedly Jointly Strongly Connected Sequences are Sufficient}\label{U}

In this section we shall prove Theorem \ref{mainer}. In other words we will show that repeatedly jointly strongly connected sequences of graphs are sufficient for exponential convergence of the $x_i$ to a solution to $Ax=b$. We will do this in two parts.
 First we will consider the special case   when $Ax=b$  has a unique solution.  This case  is exactly when $\cap_{i=1}^m \ker A_i = 0$.
Since $\ker A_i  = \scr{P}_i,\;i\in\mathbf{m}$,  the uniqueness assumption is equivalent to the condition
\eq{\bigcap_{i=1}^m\scr{P}_i = 0.\label{assmp}}

Assuming $Ax=b$ has a unique solution, our goal is to derive conditions under which  $e\rightarrow 0$ since, in view of \rep{redefine}, this will imply that all $x_i$
 approach the desired solution $x^*$ in the limit at $t\rightarrow\infty$.  To accomplish this it is clearly   enough to prove that the  matrix product $(P(F(t)\otimes I)P)\dots (P(F(2)\otimes I)P)(P(F(1)\otimes I)P)$ converges to the zero matrix exponentially fast  under the hypothesis of
    Theorem \ref{mainer}. Convergence of such matrix products is an immediate consequence of the main technical result of this paper, namely Theorem \ref{nmeets},  which we provide below.

 To state Theorem \ref{nmeets}, we need a way to quantify the sizes of matrix products of the form
 $(P(F(t)\otimes I)P)\dots (P(F(2)\otimes I)P)(P(F(1)\otimes I)P)$.
  For this purpose  we introduce a somewhat unusual but very useful concept, namely a
   special ``mixed-matrix'' norm:
Let  $|\cdot|_2$ and
$|\cdot |_{\infty}$
 denote   the standard induced
  two norm and
   infinity  norm  respectively
    and write
 $\R^{mn\times mn}$ for  the vector space of all $m\times m$  block matrices $Q = \matt{Q_{ij}}$
whose $ij$th entry     is a matrix $Q_{ij}\in\R^{n\times n}$.
  We define the {\em mixed matrix norm } of $Q\in\R^{mn\times mn}$, written $||Q||$, to be
\eq{||Q|| = |\langle Q\rangle |_{\infty}\label{mmn}}
 where $\langle Q\rangle $ is the  matrix in $\R^{m\times m}$  whose $ij$th entry is $|Q_{ij}|_2$.
It is very easy to verify that $||\cdot ||$ is in fact a norm. It is even sub-multiplicative \{cf. Lemma \ref{sub}\}.

To state Theorem \ref{nmeets}, we also need the following idea. Let $l$ be a positive integer.
A compact subset $\scr{C}$ of $m\times m$ stochastic matrices
with graphs in $\scr{G}_{sa}$ is {\em l-compact}
if  the set  $\scr{C}_l$ consisting
 of all sequences
$S_1,S_2,\ldots, S_l, \;S_i\in\scr{C},$ for which the graph $\gamma(S_lS_{l-1}\cdots S_1)$ is strongly connected,
 is nonempty and compact. Thus any nonempty compact subset of  $m\times m$
 stochastic matrices with strongly connected graphs  in $\scr{G}_{sa}$ is
 $1$-compact. Some examples
   of compact subsets which are $l$-compact  are discussed on  page 595 of \cite{reachingp1}.

The key technical result we will need   is as follows.

\begin{theorem}Suppose that \rep{assmp} holds. Let $l$ be a positive integer.
 Let $\scr{C}$ be an $l$-compact subset of  $m\times m$ stochastic matrices
and define
$$\lambda  =   (\sup_{\scr{H}_{\omega}\in\scr{C}_l}\;\sup_{\scr{H}_{\omega-1}\in\scr{C}_l},
 \cdots \;\sup_{\scr{H}_{1}\in\scr{C}_{l}}
||P(Q_{\omega l}\otimes I) P(Q_{\omega l-1}\otimes
 I)   \cdots \left. P(Q_1\otimes I)P|| \right )^{\frac{1}{\omega l}}$$
\noindent where $\omega = (m-1)^2$ and for  $i\in\{1,2,\ldots, \omega\}$,
$\scr{H}_i$  is the subsequence $
 Q_{(i-1)l +1},Q_{(i-1)l +2},\ldots, Q_{il}$.   Then $\lambda <1$, and
 for any infinite sequence
of stochastic matrices $S_1,S_2,\ldots $ in $\scr{C}$ whose graphs form a
sequence $\gamma(S_1),\gamma(S_2),\ldots $ which is repeatedly
jointly strongly connected by contiguous subsequences of length $l$,
 the following inequality holds.
\eq{||P(S_t\otimes I)P(S_{t-1}\otimes I)\cdots  P(S_1\otimes I)P||\leq \lambda^{(t-l\omega)}.\label{limits}}
\label{nmeets}\end{theorem}

The ideas upon which Theorem \ref{nmeets} depends is actually pretty simple.  One breaks the infinite product
$$\cdots P(S_t\otimes I)P(S_{t-1}\otimes I)\cdots  P(S_1\otimes I)P$$ into contiguous sub-products
$P(S_{kl}\otimes I)P(S_{{kl}-1}\otimes I)\cdots  P(S_k\otimes I)P,\;k\geq 1$ of length $l$ with $l$ chosen long enough  so that each sub-product is a contraction in the mixed matrix norm \{Proposition \ref{ppo}\}. Then using the sub-multiplicative property of the mixed matrix norm \{Lemma \ref{sub}\}, one  immediately obtains \rep{limits}.
This theorem will be proved in \S\ref{tr}.

Next we will consider the
 general case in which \rep{assmp} is not presumed to hold. This is the case when $Ax=b$
does not have a unique solution. We will deal with this case in several steps.
First  we will \{in effect\}  ``quotient out'' the subspace $\cap_{i=1}^m\scr{P}_i$ thereby obtaining
a  subsystem  to which Theorem \ref{nmeets}  can be applied. Second we will show that the part of the system state we didn't consider in the first step,
  satisfies  a standard unconstrained consensus update equation to which  well known convergence results
   can be directly applied.
The first step makes use of  the following lemma.

\begin{lemma}Let $Q'$ be any matrix whose columns form an orthonormal basis for the orthogonal complement of the
subspace $\cap_{i=1}^m\scr{P}_i$ and define $\bar{P}_i = QP_iQ',\;i\in\mathbf{m}$. Then the following
statements are true.

1. Each $\bar{P}_i,\;i\in\mathbf{m},$ is an orthogonal projection matrix.

2. Each $\bar{P}_i,\;i\in\mathbf{m},$  satisfies $QP_i = \bar{P}_i Q$.

 3. $\bigcap_{i=1}^m \bar{\scr{P}}_i = 0$.

\label{vlad}\end{lemma}

\noindent{\bf Proof of Lemma \ref{vlad}:} Note that $\bar{P}_i^2 =QP_iQ'QP_iQ'=QP_i^2Q' =QP_iQ' = \bar{P}_i,\;i\in\mathbf{m}$, so
each $\bar{P}_i$ is idempotent; since each $\bar{P}_i$ is clearly symmetric, each must be an orthogonal projection matrix.
Thus property 1 is true.

Since $\ker Q = \cap_{i=1}^m \scr{P}_i$, it must be true that  $\ker Q \subset  \scr{P}_i,\;i\in\mathbf{m}$. Thus
$P_i\ker Q =\ker Q,\;i\in\mathbf{m}$.  Therefore  $QP_i\ker Q = 0$ so $\ker Q\subset \ker QP_i$.  This plus the fact that $Q$ has
 linearly independent rows means that the equation $QP_i = XQ$ has a unique solution $X$. Clearly $X = QP_iQ'$,
 so $X=\bar{P}_i$.  Therefore property 2 is true.

 Pick $x\in\cap_{i=1}^m\bar{\scr{P}}_i$.  Then $x\in\bar{\scr{P}}_i,\;i\in\mathbf{m},$ so
 there exist $w_i$ such that $x_=\bar{P}_iw_i,\;i\in\mathbf{m}$.  Set   $y=Q'x$ in which case $x=Qy$; thus  $y
  = Q'\bar{P}_iw_i,\;i\in\mathbf{m} $. In view of property 2 of Lemma \ref{vlad},
 $y= P_iQ'w_i,\;i\in\mathbf{m}$  so $y\in\cap_{i=1}^m\scr{P}_i$.  Thus $Qy = 0$.  But $x=Qy$ so $x=0$.  Therefore
property 3 of Lemma \ref{vlad} is true.
$\qed $

\noindent{\bf Proof of Theorem  \ref{mainer}:} Consider first the case when $Ax = b$ has a unique solution. Thus the hypothesis of Theorem \ref{nmeets} that   \rep{assmp} hold, is satisfied. Next observe that since directed graphs
 in $\mathbb{G}_{sa}$ are bijectively related to flocking matrices,
 the set $\scr{F}_l $ of distinct subsequences
 $F((k-1)l +1),F((k-1)l+2),\ldots, F(kl),\;k\geq 1$,   encountered along any trajectory of \rep{pup} must be a finite
and thus compact set.  Moreover for some finite integer $\tau_0\geq 0$,  the composed graphs $\gamma(F(kl))\circ\gamma(F(lk-1)\circ\cdots F(l(k-1) +1)),
\;\;  k\geq \tau_0$  must be strongly connected because the neighbor graph
 sequence $\mathbb{N}(t),\;t\geq 1$ is repeatedly jointly strongly connected by subsequences of length $l$
 and    $\gamma(F(t)) = \mathbb{N}(t),\;t\geq 1$. Hence Theorem \ref{nmeets} is applicable to the matrix product $(P(F(t)\otimes I)P)\dots (P(F(2)\otimes I)P)(P(F(1)\otimes I)P)$. Therefore  for suitably defined nonnegative  $\lambda <1 $, this product converges to the zero matrix as fast as $\lambda^t$ converges to $0$. This and \rep{pup} imply that
$e(t)$ converges to zero just as fast.  From this and \rep{redefine} it follows that each $x_i(t)$ converges
exponentially fast to $x^*$.  Therefore  Theorem \ref{mainer} is
 true for the case when $Ax=b$ has
 a unique solution.

Now consider the case when $Ax = b$ has more than one solution. 
Note that property 2 of Lemma \ref{vlad} implies that $QP_iP_j = \bar{P}_i\bar{P}_jQ$ for all $i,j\in\mathbf{m}$.  Thus if
we define $\bar{e}_i = Qe_i,\;i\in\mathbf{m}$,   then from \rep{na1}
\eq{\bar{e}_i(t+1) =
\frac{1}{m_i(t)}\bar{P}_i\sum_{j\in\scr{N}_i(t)}\bar{P}_j\bar{e}_j(t),\;\;t\geq 1,\;\;i\in\mathbf{m}.\label{nna1}}
 Observe that \rep{nna1} has exactly the same form as \rep{na1} except for the  $\bar{P}_i$ which replace the $P_i$.
But in view of Lemma \ref{vlad},  the $\bar{P}_i$ are also orthogonal  projection matrices  and $\cap_{i=1}^m\bar{\scr{P}}_i = 0$.
Thus Theorem \ref{nmeets}
  is also applicable to the system of iterations \rep{nna1}. Therefore
 $\bar{e}_i \rightarrow 0$ exponentially
  fast as $t\rightarrow\infty$. 

To deal with what is left,
define $z_i = e_i-Q'\bar{e}_i,\;i\in\mathbf{m}$.
Note that $Qz_i = Qe_i - \bar{e}_i$ so $Qz_i = 0,i\in\mathbf{m}$.  Thus $z_i(t)\in\cap_{j=1}^m \scr{P}_j,\;i\in\mathbf{m}$.
 Clearly $P_jz_i(t) = z_i(t),\;i,j\in\mathbf{m}$.
Moreover  from property 2 of Lemma \ref{vlad}, $P_iQ' = Q'\bar{P}_i$.  These expressions, and \rep{nna1}
imply that
\eq{z_i(t+1) =
\frac{1}{m_i(t)}\sum_{j\in\scr{N}_i(t)}z_j(t),\;\;t\geq 1,\;\;i\in\mathbf{m}.\label{dinner}}
These equations are the update equations for the standard unconstrained consensus problem treated in \cite{reachingp1} and elsewhere for case
 when the $z_i$ are scalars.  It is well known that for the scalar case, a sufficient
 condition for  all $z_i$ to converge exponentially fast to the same value is that the neighbor  graph sequence
  the $\mathbb{N}(t),\;\;t\geq 1$  be  repeatedly jointly strongly connected \cite{reachingp1}.
But since   the  vector update  \rep{dinner} decouples into
 $n$ independent scalar update equations,  the convergence conditions for the scalar equations apply  without change
 to the vector case  as well. Thus  all $z_i$ converge  exponentially  fast
  to the same limit  in $z^*\in\bigcap_{i = 1}^m\scr{P}_i$. So do all of the $e_i$
  since $e_i =z_i+Q'\bar{e}_i,\;i\in\mathbf{m}$, and all $\bar{e}_i $ converge to zero exponentially fast. Therefore all
$x_i$ defined by  \rep{a1} converge to the same limit $x^* + z^*$ which solves $Ax=b$.
This concludes the  proof of Theorem \ref{mainer} for the case when $Ax=b$ does not have a unique solution. $\qed $

\subsection{Repeatedly Jointly Strongly Connected Sequences are Necessary}


 \label{nec} In this section we shall explain why  the  of exponential convergence of the $x_i(t)$ to a solution can only occur if the sequence of neighbor graphs $\mathbb{N}(t),\;t\geq 0$ referred to in the statement of  Theorem \ref{mainer.nes}, is repeatedly jointly strongly connected. To do this we need the following concepts  from \cite{seneta} and \cite{gallager}.
A vertex $j$ of a directed graph $\mathbb{G}$ is said to be {\em reachable} from vertex $i$ if either $i=j$ or
 there is a directed path from $i$ to $j$. Vertex  $i$ is called  {\em essential} if it is reachable from all vertices which
are reachable from $i$. It is known that every directed graph has at least one essential vertex \{Lemma 10 of
\cite{deterministic}\}.

Vertices $i$ and $j$ in $\mathbb{G}$ are called {\em mutually reachable} if each is reachable from the other.
Mutual reachability is an equivalence relation on $\mathbf{m}$.  Observe that if $i$ is an essential vertex in $\mathbb{G}$,
 then every vertex in the equivalence class of $i$ is essential. Thus each directed graph possesses at least
  one mutually reachable equivalence class whose members are all
  essential.  Note also that a strongly connected graph has exactly one mutually reachable equivalence class.\vspace{.1in}

\noindent{\bf Proof of Theorem \ref{mainer.nes}:}
Consider first the case when $Ax = b$ has a unique solution.  In this case, the unique equilibrium of  \rep{pup} at the origin
must be exponentially stable.
 Since  exponential stability and uniform asymptotic stability  are equivalent properties for linear systems, it is enough to show that uniform asymptotic stability of \rep{pup}  implies that  the sequence of neighbor graphs $\mathbb{N}(t),\;t\geq 0$ is repeatedly jointly strongly connected. Suppose therefore that \rep{pup} is a uniformly asymptotically  stable system.

To show that repeatedly jointly strongly connected sequences are necessary for uniform asymptotic stability, we suppose the contrary; i.e. suppose that  $\mathbb{N}(1),\mathbb{N}(2),\ldots $ is not
a repeatedly jointly strongly connected sequence. Under these conditions, we claim that
 for every pair of positive integers $l$ and $\tau $, there  is an integer $k> \tau $ such that the composed graph
$\mathbb{N}(k+l-1)\circ\cdots\circ \mathbb{N}(k+1)\circ\mathbb{N}(k)$
is not strongly connected. To justify this claim, suppose that  for some pair $(l,\tau )$, no such   $k$  exists;  thus for this pair, the graphs $\mathbb{N}(p+l-1)\circ\cdots\circ \mathbb{N}(p+1)\circ\mathbb{N}(p),\;\;p\geq \tau $ are all
strongly connected so the sequence $\mathbb{N}(1)$, $\mathbb{N}(2),\ldots $ must be repeatedly jointly strongly connected.
 But this contradicts the hypothesis that   $\mathbb{N}(t),\;t\geq 0$ is not
a repeatedly jointly strongly connected sequence. Therefore for any pair of positive integers $l$ and $\tau $ there  is an integer $k>\tau$ such that the composed graph $\mathbb{N}(k+l-1)\circ\cdots\circ \mathbb{N}(k+1)\circ\mathbb{N}(k)$
is not strongly connected.

Let $\Phi(t,\tau )$ be the state transition matrix of $P(F(t)\otimes I)P$.
Since \rep{pup} is uniformly asymptotically stable, for each real number $\epsilon >0$ there exist  positive integers
$t_{\epsilon}$ and $T_{\epsilon}$ such that $||\Phi(t+T_{\epsilon},t )||<\epsilon $ for all $t>t_{\epsilon}$.
Set $\epsilon = 1$ and let $t_1$ and $T_1$ be any pair of such integers.  Since $\mathbb{N}(1),\mathbb{N}(2),\ldots $ is not a repeatedly strongly connected sequence,  there must be an integer $t_2 >t_1$  for which the composed graph
$$\mathbb{G} = \mathbb{N}(t_2+T_1-1)\circ\cdots\circ \mathbb{N}(t_2+1)\circ\mathbb{N}(t_2)$$
is not strongly connected. Since $t_2>t_1$, the hypothesis of uniform asymptotic stability ensures that
\eq{||\Phi(t_2 +T_1,t_2)||<1.\label{pty}}

In view of the discussion just before the proof of Theorem \ref{mainer.nes},  $\mathbb{G}$ must have at least one
mutually reachable equivalence class $\scr{E}$ whose members are all essential.
Note that if $\scr{E} $ where equal to $ \mathbf{m}$, then $\mathbb{G}$ would have to be strongly connected. But $\mathbb{G}$ is not strongly connected so $\scr{E}$ must be a strictly proper subset of $\mathbf{m}$ with $k<m$ elements.
Suppose that $\scr{E} = \{v_1,v_2,\ldots,v_k\}$ and let $\bar{\scr{E}} = \{v_{k+1},\ldots,v_m\}$ be the complement of
$\scr{E}$ in $\mathbf{m}$. Since every vertex in $\scr{E}$ is essential, there are no arcs in $\mathbb{G}$  from
$\scr{E}$ to $\bar{\scr{E}}$. But the arcs of  each $\mathbb{N}(t),\;t\in \{t_2,t_2+1,\ldots t_2+T_1-1\}$ must all be arcs in $\mathbb{G}$ because
each $\mathbb{N}(t)$ has self-arcs at all vertices. Therefore  there cannot be an arc from $\scr{E}$ to $\bar{\scr{E}}$
in any $\mathbb{N}(t),\;t\in\{t_2,t_2+1,\ldots t_2+T_1-1\}$.

Let $\pi$ be a permutation on $\mathbf{m}$ for which $\pi (v_j) = j,\;j\in\mathbf{m}$  and let $Q$ be the corresponding permutation matrix. Then for $t\in\{t_2,t_2+1,\ldots t_2+T_1-1\}$,  the transformation $F(t)\longmapsto QF(t)Q'$ block triangularizes $F(t)$. Set  $\bar{Q} = Q\otimes I$. Note that  $\bar{Q}$ is  a permutation matrix and  that $\bar{Q}P\bar{Q}'$ is a block diagonal, orthogonal projection matrix whose $j$th diagonal block is $P_{\pi(v_j)},\;\;j\in\mathbf{m}$. Because each  $QF(t)Q'$ is block triangular, so are the matrices
$\bar{Q}P(F(t)\otimes I)P\bar{Q}',\;t\in\{t_2,t_2+1,\ldots t_2+T_1-1\}$.
Thus for  $t\in\{t_2,t_2+1,\ldots t_2+T_1-1\}$, there are matrices  $A(t),B(t)$ and $C(t)$ such that
$$\bar{Q}P(F(t)\otimes I)P\bar{Q}' = \matt{A(t) & B(t)\cr 0 &C(t)}.$$

 Let $k$ be the number of elements in $\scr{E}$. For $t\in\{t_2,t_2+1,\ldots t_2+T_1-1\}$, let $\bar{S}(t)$ be that $(m-k)\times (m-k)$  submatrix of $F(t)$ whose $ij$th entry is
 the $v_{i+k}v_{j+k}$th entry  of $F(t)$, for all and $i,j\in\{1,2,\ldots,m-k\}$.  In other words, $\bar{S}(t)$ is that submatrix of $F(t)$ obtained by deleting rows and columns whose indices are in $\scr{E}$. Since each $F(t),\;t\in\{t_2,t_2+1,\ldots t_2+T_1-1\}$ is a stochastic matrix and there are no arcs from $\scr{E}$ to $\bar{\scr{E}}$, each corresponding $\bar{S}(t)$ is a stochastic matrix as well. Set $\bar{P} = {\rm block\;diagonal}
 \{P_{v_{k+1}},P_{v_{k+2}},\ldots, P_{v_m}\}$ in which case $C(t) = \bar{P}(\bar{S}(t)\otimes I)\bar{P}$.
 Since $\{P_1,P_2,\ldots ,P_m\}$ is a non-redundant  set and $\bar{\scr{E}}$ is a strictly proper subset of $\mathbf{m}$, $\bigcap_{i\in\bar{\scr{E}}}\scr{P}_i\neq 0$. Let $z$ be any nonzero vector in $\bigcap_{i\in\bar{\scr{E}}}\scr{P}_i$.
 in which case $P_iz=z,\;i\in\bar{\scr{E}}$. Then $C(t)\bar{z} = \bar{P}(\bar{S}(t)\otimes I)\bar{P}\bar{z}$ where $\bar{z} = \matt{z' &z'&\cdots &z'}'_{(m-k)n\times 1}$.  Note that
$$\bar{Q}\Phi(t_2+T_1,t_2)\bar{Q}' = (\bar{Q}P(F(t_2+T_1-1)\otimes I)P\bar{Q}')\cdots (\bar{Q}P(F(t_2)\otimes I)P\bar{Q}') = \matt{A &B\cr 0 &C}$$
where $C=C(t_2+T_1-1)\cdots C(t_2)$. Therefore $C\bar{z} = \bar{z}$ for $C$ has an eigenvalue at $1$. Thus the state transition matrix $\Phi(t_2+T_1-1,t_2)$ has an eigenvalue at $1$ so $||\Phi(t_2+T_1-1,t_2)|| = 1$. But this  contradicts \rep{pty}.  It follows that  the sequence $\mathbb{N}(1),\mathbb{N}(2),\ldots $  must be repeatedly jointly strongly connected if $Ax=b$ has a unique solution.

We now turn to the general case in which $Ax=b$ has more than one solution.  Since by assumption, $A\neq 0$,
 the matrix $Q$ defined in the statement  of Lemma \ref{vlad} is not the zero matrix and so the subsystem defined by \rep{nna1} has positive state space dimension.  Moreover, exponential convergence of the overall system implies that  this subsystem's unique equilibrium at the   origin is exponentially stable. Thus the preceding arguments apply so
 the sequence of neighbor graphs must be repeatedly jointly strongly connected in this case too.  $\qed $

\subsection{Justification for Theorem \ref{nmeets}}\label{gal}

In this section we develop the ideas needed to prove Theorem \ref{nmeets}.  We begin with the  following
lemma which provides several elementary but useful facts  about orthogonal projection matrices.
\begin{lemma} For any nonempty set of $n\times n$ real orthogonal projection matrices  $\{P_1,P_2,\ldots, P_k\}$
 \eq{|P_kP_{k-1}\cdots P_1|_2\leq 1.\label{snow1}}
Moreover, \eq{|P_kP_{k-1}\cdots P_1|_2< 1\label{snow2}}  if and only
 if \eq{\bigcap_{i=1}^k\scr{P}_i = 0.\label{snow3}}\label{crack}\end{lemma}

\noindent{\bf Proof of Lemma \ref{crack}:} To avoid cumbersome notation,  throughout this proof  we drop
the subscript $2$ and write $|\cdot|$ for $|\cdot |_2$.
 To establish \rep{snow1}, We make use of the fact that the eigenvalues of any projection matrix are either $0$ or $1$.
 But the projection matrices of interest here are orthogonal and thus symmetric.  Therefore each singular value  of
 each $P_i$ must be either $0$ or $1$.  It follows that
 $|P_i|\leq 1,\;i\in\mathbf{k}$.
  The inequality in \rep{snow1} follows at once  the fact
that $|\cdot|$ is sub-multiplicative.

To prove the equivalence of \rep{snow2} and \rep{snow3}
 suppose  first that \rep{snow2} holds.  Let $x$ be any vector in $\bigcap_{i=1}^k\scr{P}_i$.
  Then $P_kP_{k-1}\cdots P_1x = x$.
 Since \rep{snow2} holds, $P_kP_{k-1}\cdots P_1$  must be a discrete time stability matrix. Therefore  $P_kP_{k-1}\cdots P_1$
 cannot have an eigenvalue at $1$
so $x=0.$  It follows that \rep{snow3} is true.

To proceed we will first need to justify the following claim:
If $\{Q_{1},Q_{2}, \ldots ,Q_s\}$ is any nonempty subset of
  $s\leq m$  projection matrices from $\{P_1,P_2,\ldots, P_k\}$ and
  $x\in\R^n$ is any vector for which
$|Q_1\cdots Q_{s-1}Q_sx| = |x|$, then $Q_ix = x,\;i\in\{1,2,\ldots,s\}$. To prove this claim,
 suppose first that
 $Q\in\{P_1,P_2,\ldots, P_k\}$ and that $|Qx|=|x|$ for some  $x\in\R^n$.
 Write $x=y+z$ where  $y\in\scr{Q}$ and $z\in\scr{Q}^{\perp}$. Then $Qx = y$ so $|y|=|x|$.  But
 $|y|^2+|z|^2 = |x|^2$ so $z=0$.  Therefore $Qx=x$ so the claim is true for $s=1$.

Now  fix $q<k$ and  suppose  that the claim is true for every
value of $s\leq q$. Let $x$ be a  vector for which
$|Q_{1}\cdots Q_{q} Q_{q+1}x| = |x|$.  Then
$|x| = |Q_{1}\cdots Q_{q} Q_{q+1}x|\leq |Q_{q+1}x|\leq |x|$
because $|\cdot|$ is sub-multiplicative and because \rep{snow1} holds for any nonempty set of projection matrices.
Clearly $|Q_{q+1}x| = |x|$; therefore  $Q_{q+1}x = x$ because the claim is
 true for single projection matrices. Therefore $Q_{1}\cdots Q_{q} Q_{q+1}x=Q_{1}\cdots Q_{q}x $
 so $|Q_{1}\cdots Q_{q}x |   = |x|$.  From this and the inductive hypothesis it follows that
$Q_ix = x,\;i\in\{1,2,\ldots,q\}$.  Thus the claim is true for all $s\leq q+1$.
 It follows by induction that the claim is true.

To complete the proof, suppose that \rep{snow3} holds and let $x$ be any vector for which
$|P_kP_{k-1}\cdots P_1x| = |x|$. In view of the preceding claim, $P_ix=x,\;i\in\{1,2,\ldots,k\}$.
This implies that $x\in\cap_{1}^k\scr{P}_i$, and thus because of \rep{snow3} that $x=0$.  Thus
$P_kP_{k-1}\cdots P_1$ cannot have a singular value at $1$. This and \rep{snow1} imply that
  \rep{snow2} is true.  $\qed $


\subsubsection{Projection Matrix Polynomials} \label{PRO}

To proceed we need to develop a language for talking about matrix products of the form
$(P(S_q\otimes I)P)\dots (P(S_2\otimes I)P)(P(S_1\otimes I)P)$ where the $S_i$ are  $m\times m$ stochastic matrices. Such matrices  can be viewed as  partitioned matrices  whose $m^2$  blocks are specially structured  $n\times n$ matrices.  We begin by introducing some concepts  appropriate to the individual blocks.

Let $\{P_1,P_2,\ldots , P_m\}$ be a set of $n\times n$ orthogonal projection matrices. 
We will be interested
in  matrices of the form
\eq{\mu(P_1,P_2,P_3,\ldots ,P_m) = \sum_{i=1}^{d}\lambda_i P_{h_i(1)}P_{h_i(2)}\cdots P_{h_i(q_i)}\label{search}}
where $q_i$ and $d$ are positive integers, $\lambda_i $ is a real positive
   number, and for each $j\in\{1,2,\ldots,q_i\}$,
 $h_i(j)$  is an integer in $\{1,2,\ldots,m\}$.
 We call such matrices together
 with the $n\times n$ zero matrix, {\em projection matrix polynomials}.
  In the event $\mu$ is nonzero,  we refer to  the $\lambda_i$  as the {\em coefficients}  of $\mu$.
 Note that each $n\times n$
 block of any partitioned  matrix of the form $(P(S_q\otimes I)P)\dots (P(S_2\otimes I)P)(P(S_1\otimes I)P)$ is a projection matrix polynomial.
   The set of  projection matrix polynomials, written $\mathbb{P}$,
    is clearly closed under matrix addition and multiplication.
Let us note from the triangle inequality, that
$$|\mu(P_1,P_2,P_3,\ldots ,P_m)|_2\leq
\sum_{i=1}^{d}\lambda_i|P_{h_i(1)}P_{h_i(2)}\cdots P_{h_i(q_i)}|_2.$$
From this and \rep{snow1} it follows that
\eq{|\mu(P_1,P_2,P_3,\ldots ,P_m)|_2\leq \lceil\mu(P_1,P_2,P_3,\ldots ,P_m)\rceil\label{gg}}
where
$\lceil\mu(P_1,P_2,P_3,\ldots ,P_m)\rceil =\sum_{i=1}^{d}\lambda_i$ if $\mu \neq 0$ and $\lceil \mu \rceil = 0$ if $\mu=0$.
We call $\lceil\mu\rceil$ the {\em nominal bound} of $ \mu$.
Notice that the actual $2$ norm of $\mu$  will be strictly less than its nominal bound provided
at least one ``component'' of $\mu$  has    a $2$ norm less than
one where by a {\em component} of $\mu$ we mean  any matrix product
    $P_{h_i(1)}P_{h_i(2)}\cdots P_{h_i(q_i)}$ appearing in the sum
in \rep{search} which defines $\mu$.   In view of Lemma \ref{crack}, a sufficient
condition for $P_{h_i(1)}P_{h_i(2)}\cdots P_{h_i(q_i)}$
 to have a $2$ norm
less than $1$ is that
$$\bigcap_{j=1}^{q_i} {\rm Im}(P_{h_i(j)}) = 0.$$
Thus if $\bigcap_{i=1}^m\scr{P}_i = 0$, this in turn will always be true if  each of the projections matrices
in the set
 $\{P_1,P_2,\ldots,P_m\}$
appears in the component  $P_{h_i(1)}P_{h_i(2)}\cdots P_{h_i(q_i)}$ at least once.
Prompted by this we say that a  nonzero projection matrix polynomial
 $\mu(P_1,P_2,P_3,\ldots ,P_m)$  is {\em complete}  if  it has a    component
   $P_{h_i(1)}P_{h_i(2)}\cdots P_{h_i(q_i)}$ within which each of the
   projections matrices $P_j,\;j\in\{1,2,\ldots,m\}$
appears at least once. Assuming  $\bigcap_{i=1}^m\scr{P}_i = 0$,
 complete  projection matrix
   polynomials are thus a class of projection matrix polynomials
    with $2$-norms strictly less than their nominal bounding values.
    The converse of course is not necessarily so.

\subsubsection{Projection Block Matrices}\label{PBM}

The ideas just discussed extend in an natural way to ``projection block matrices.''  By an $m\times m$
 {\em projection block matrix}  is meant a block  partitioned matrix of the form
$$M = \matt{\mu_{ij}(P_1,P_2,\ldots,P_m)}_{m\times m}.$$
An $m\times m$  projection block matrix is thus an $nm\times nm$ matrix of
real numbers partitioned into $n\times n$  sub-matrices which are projection matrix polynomials.
The set of all $m\times m$ projection block matrices, written $\mathbb{P}^{m\times m}$, is clearly closed under multiplication. Note that any matrix of the form $(P(S_q\otimes I)P)\dots (P(S_2\otimes I)P)(P(S_1\otimes I)P)$ is a projection block matrix.

By the {\em nominal bound of } $M=$ $\matt{\mu_{ij}(P_1,P_2,\ldots,P_m)}_{m\times m}\in\mathbb{P}^{m\times m}$, written $\lceil M\rceil $, is meant the $m\times m$ matrix whose $ij$th entry is the
nominal bound of $\mu_{ij}(P_1,P_2,\ldots,P_m)$.  Using \rep{gg} it is quite easy to verify that
\eq{\langle M \rangle\leq \lceil M\rceil \label{pp1}}
where the inequality
is intended entry-wise.
The definition of nominal bound of a projection matrix polynomial implies that for all $\mu_1,\mu_2\in\mathbb{P}$,
$\lceil \mu_1\mu_2\rceil = \lceil \mu_1\rceil\lceil\mu_2\rceil  $ and
$\lceil \mu_1+\mu_2\rceil = \lceil \mu_1\rceil+\lceil\mu_2\rceil  $.
From this it follows that
\eq{\lceil M_1M_2\rceil = \lceil M_1\rceil\lceil M_2\rceil,\;\;\;M_1,M_2\in\mathbb{P}^{m\times m}.\label{trouble}}

In order to measure the sizes of matrices in $\mathbb{P}^{m\times m}$ we shall make
 use of the  mixed matrix norm $||\cdot||$ defined earlier in \rep{mmn}.  A critical property of this norm is that it is
 sub-multiplicative:

\begin{lemma}
$$||AB||\leq ||A||||B||,\;\;\;\;\forall A,B\in\R^{mn\times mn}.$$
\label{sub}\end{lemma}

\noindent{\bf Proof of Lemma \ref{sub}:}
Note first that
$$\langle AB \rangle = \left [\sum_{k=1}^n| A_{ik}B_{kj}|_2 \right ]_{m\times m}.$$
But $| A_{ik}B_{kj}|_2 \leq | A_{ik}|_2| B_{kj}|_2 $  so
$$\sum_{k=1}^m| A_{ik}B_{kj}|_2\leq \sum_{k=1}^m| A_{ik}|_2| B_{kj}|_2 =
\matt{|A_{i1}|_2 &| A_{i2}|_2 &\cdots &| A_{im}|_2 } \matt{
| B_{1j}|_2 \cr | B_{2j}|_2\cr \vdots  \cr | B_{mj}|_2 }.   $$
Clearly
$ \langle AB \rangle \leq \langle A\rangle\langle B \rangle$.
 It follows from this and the fact that the infinity norm is sub-multiplicative that
$ | \langle AB \rangle |_{\infty} \leq | \langle A\rangle|_{\infty}|\langle B \rangle |_{\infty} $
Thus the lemma is true. $\qed $

It is worth noting that the preceding properties of $||\cdot ||$ remain true for
 any pair of  standard matrix norms
 provided both are sub-multiplicative. It is conceivable  that the mixed matrix norm  which results when the $1$ -norm  is used in place of
  the $2$-norm,
 will find application in the study of distributed compressed sensing algorithms \cite{compsense}.
   The notion of a mixed matrix norm has been used before  although references to the subject are
somewhat obscure.

Let $M = \matt{\mu_{ij}}_{m\times m}$ be a matrix in $\mathbb{P}^{m\times m}$.
 Since $\langle M\rangle = \matt{ |\mu_{ij}|_2}_{m\times m}$,            it is possible to rewrite  \rep{pp1} as
\eq{\langle M \rangle \leq \lceil M\rceil,\;\;\;M\in\mathbb{P}^{m\times m}.\label{food}} Therefore
\eq{||M|| \leq |\lceil M \rceil |_{\infty},\;\;\;M\in\mathbb{P}^{m\times m}.\label{p5}}
Thus in the case when $\lceil M \rceil$ turns out to be a stochastic matrix,
$||M||\leq 1$.   In other words, when $\lceil M \rceil$
is a stochastic matrix,  $M$ is non-expansive. As will soon become clear, this is exactly the case we are interested in.

What we are especially interested in are conditions under which $M$ is a contraction in the mixed matrix norm
we have been discussing under the assumption that $\bigcap_{i=1}^m\scr{P}_i = 0$.
 Towards this end, let us note first that the sum of the terms in any given
row $i$ of $\langle M\rangle $ will be strictly
less than the sum of the terms in row $i$ of $\lceil M\rceil $ provided at least
 one sub-matrix $\mu_{ij}$  in block row $i$ of $M$ is complete. 
It follows at once
that the inequality in \rep{p5} will be strict if every row of $M$ has this property.
 We have proved the following
proposition.


\begin{proposition}   Any matrix $M$ in $\mathbb{P}^{m\times m}$  whose nominal bound is stochastic,
is non-expansive
in the mixed matrix norm. If, in addition,  $\bigcap_{i=1}^m\scr{P}_i = 0$ and at least one entry in each block row of $M$ is complete, then
 $M$ is a contraction in the mixed matrix norm.
\label{ve}\end{proposition}

\subsubsection{Technical Results} \label{tr}

We now return to the study of  matrix products  of the form
$P(S_q\otimes I)P(S_{q-1}\otimes I)\cdots  P(S_1\otimes I)P$
where $P = {\rm diagonal}\;\{P_1,P_2,\ldots, P_m\}$, $S_i$ is an $m\times m$  stochastic matrix,
 and $I$ is the $n\times n $ identity.
As  noted  earlier,  each such matrix product
 is a projection block matrix  in $\mathbb{P}^{m\times m}$.
Our goal is to  state a sufficient condition under which
 any such matrix product is a contraction in the mixed matrix norm.
To do this let us note first that
\eq{\lceil P(S_q\otimes I)P(S_{q-1}\otimes I)\cdots  P(S_1\otimes I)P\rceil = S_qS_{q-1}\cdots S_1\label{trouble2}}
because of \rep{trouble} and the fact that $\lceil P(S\otimes I)P \rceil = S$ for any  stochastic matrix $S$.
Thus in view of Proposition \ref{ve},   $P(S_q\otimes I)P(S_{q-1}\otimes I)\cdots  P(S_1\otimes I)P$ will be a contraction
assuming \rep{assmp} holds, if each of its block rows contains an entry which is complete.

 To proceed  we need to generalize
 the idea of a  repeated jointly strongly connected sequence to sequences of finite length.
A finite sequence of graphs $\mathbb{G}_1, \mathbb{G}_2,\ldots \mathbb{G}_l$  in $\scr{G}_{sa}$
is {\em $l$-connected}  if the composed graph  $\mathbb{G}_l\circ \mathbb{G}_{l-1}\circ\cdots\circ \mathbb{G}_1$
 is strongly  connected.  More generally, finite sequence
   $\mathbb{G}_1, \mathbb{G}_2,\ldots \mathbb{G}_p$ is
 {\em repeatedly $l$-connected} for some positive integer $l$, if each of the  composed graphs
$\mathbb{H}_k=\mathbb{G}_{kl}\circ\mathbb{G}_{kl-1}\circ\cdots\circ\mathbb{G}_{(k-1)l+1},\; k\in \mathbf{q},$
is strongly connected; here  $q$ is the unique integer quotient of $p$ divided by $l$. Note that  if
$\mathbb{G}_1, \mathbb{G}_2,\ldots \mathbb{G}_p$ is such a sequence,   the composed graph $\mathbb{H} =
\mathbb{G}_{p}\circ \mathbb{G}_{p-1}\circ\cdots\circ \mathbb{G}_{l(q-1)+1}$ is also strongly connected
because $\mathbb{H} =\mathbb{G}_{p}\circ \mathbb{G}_{p-1}\circ \cdots\circ\mathbb{G}_{ql+1}\circ\mathbb{H}_q$
and because in $\scr{G}_{sa}$, the arc sets  of any two graphs are   contained in the arc set of their composition.




\begin{proposition} Suppose that \rep{assmp} holds. Let $S_1,S_2,\ldots S_p $  be a finite set of $m\times m$ stochastic matrices
whose graphs  form a sequence $\gamma(S_1),\gamma(S_2),\ldots ,\gamma(S_p)$
 which is repeatedly $l$-connected for some positive integer  $l$. If $p\geq (m-1)^2l $, then
 the matrix
$P(S_{p}\otimes I)P(S_{p}\otimes I)\cdots  P(S_1\otimes I)P$
  is a contraction in the mixed  matrix norm. \label{ppo}\end{proposition}

To prove this proposition we will make use of the following idea.
By a {\em route} over a given sequence of graphs $\mathbb{G}_1,\;\mathbb{G}_2,\ldots,\mathbb{G}_q$  in $\scr{G}_{sa}$
 is meant a sequence of vertices $i_0, i_1,\ldots, i_q$ such that for $k\in\mathbf{q}$,
  $(i_{k-1},i_k)$  is an arc in $\mathbb{G}_k$.
  A route over a sequence of graphs which are all the same graph $\mathbb{G}$, is thus a
  walk in $\mathbb{G}$.

  The definition of a route implies that if $i_0, i_1,\ldots, i_q$ is a route over
$\mathbb{G}_1,\;\mathbb{G}_2,\ldots,\mathbb{G}_q$ and  $i_q,i_{q+1},\ldots, i_p$ is a route over
$\mathbb{G}_q,\;\mathbb{G}_{q+1},\ldots,\mathbb{G}_p$, then  the `concatenated' sequence $i_0, i_1,\ldots,i_{q-1}$, $ i_q,i_{q+1},\ldots, i_p$
is a route over $\mathbb{G}_1,\;\mathbb{G}_2,\ldots,\mathbb{G}_{q-1},\;\mathbb{G}_{q},\; \mathbb{G}_{q+1},\ldots,\mathbb{G}_p$.
This clearly  remains true if more than two sequences are concatenated.

Note that the definition of composition in $\scr{G}_{sa}$  implies that if $j=i_0, i_1,\ldots, i_q=i$ is
a route over a sequence $\mathbb{G}_1,\;\mathbb{G}_2,\ldots,\mathbb{G}_q$,
then $(i,j)$ must be an arc in the composed graph $\mathbb{G}_q\circ\mathbb{G}_{q-1}\circ\cdots \circ\mathbb{G}_1$.
The definition of composition also implies the converse, namely that  if $(i,j)$ is an arc in $\mathbb{G}_q\circ\mathbb{G}_{q-1}\circ\cdots \circ\mathbb{G}_1$,
then there must exist vertices $i_1,\ldots, i_{q-1}$ for which $j=i_0, i_1,\ldots, i_q=i$ is a route over
$\mathbb{G}_1,\;\mathbb{G}_2,\ldots,\mathbb{G}_q$.

\begin{lemma} Let $S_1,S_2,\ldots S_q $ be a sequence of $m\times m$  stochastic matrices
with  graphs
$\mathbb{G}_1,\mathbb{G}_2,\ldots$, $\mathbb{G}_q$  in $\scr{G}_{sa}$ respectively. If  $j=i_0, i_1,\ldots, i_q=i$
 is a
route over the sequence
$\mathbb{G}_1,\;\mathbb{G}_2,\ldots,\mathbb{G}_q$, then the   matrix product $P_{i_q}P_{i_{q-1}}\cdots P_{i_0}$
is a component of the   $ij$th block entry of the projection block matrix
$$M=P(S_q\otimes I)P(S_{q-1}\otimes I)\cdots  P(S_1\otimes I)P.$$ \label{gum}\end{lemma}

\noindent{\bf Proof of Lemma \ref{gum}:}
First suppose $q=1$ in which case $M=P(S_1\otimes I)P$.  By definition, $(j,i)$ is an arc in $\mathbb{G}_1$; therefore $s_{ij}\neq 0$.
But the $ij$th block in $M$  is $s_{ij}P_iP_j$.  Thus the lemma is true for $q=1$.

Now suppose that  $q>1$ and that the lemma is true for all $k< q$.
Set $A=PS_qP$ and  $B=P(S_{q-1}\otimes I)P(S_{q-2}\otimes I)\cdots  P(S_1\otimes I)P$. Since $P^2=P$, $M=AB$.
Since the lemma is true for $k<q$ and  $j=i_0,i_1,i_2,\ldots, i_{q-1}$
is a route over $\mathbb{G}_1,\;\mathbb{G}_2,\ldots,\mathbb{G}_{q-1}$, the   matrix
 $P_{i_{q-1}}P_{i_{q-2}}\cdots P_{i_0}$
is a component of  the  $i_{q-1}j$th projection matrix  polynomial entry $b_{i_{q-1}j}$ of $B$.
Similarly, the matrix $P_{i_q}P{i_{q-1}}$  is  a component of the   $ii_{q-1}$th projection matrix  polynomial entry $a_{ii_{q-1}}$ of $A$.
In general, the product of any component of any nonzero projection matrix polynomial $\alpha$
with any component of any  other nonzero projection matrix polynomial $\beta$,
is a component of the product $\alpha\beta$.  It must therefore be true that
  $P_{i_q}P_{i_{q-1}}P_{i_{q-1}}P_{i_{q-2}}\cdots P_{i_0}$ is a component of the product
  $a_{ii_{q-1}}b_{i_{q-1}j}$.
But $P_{i_{q-1}}^2 = P_{i_{q-1}}$ so $P_{i_q}P_{i_{q-1}}P_{i_{q-2}}\cdots P_{i_0}$ is
 a component of  $a_{ii_{q-1}}b_{i_{q-1}j}$.
In view of the definition of matrix multiplication, the projection matrix  polynomial  $a_{ii_{q-1}} b_{i_{q-1}j}$
must appear within the  sum which defines the $ij$th block entry $\mu_{ij}$ in $M$.
Therefore
 $P_{i_q}P_{i_{q-1}}P_{i_{q-2}}\cdots P_{i_0}$ must be a component of   $\mu_{ij}$.
Thus   the lemma is true at $q$.  By induction the lemma is true for all $q>0$.$\qed $

\noindent{\bf Proof of Proposition \ref{ppo}:} Set $r=m-1$ and $\mathbb{G}_i =\gamma(S_i),\;i\in\mathbf{p}$.
  Partition the sequence $\mathbb{G}_1,\mathbb{G}_2,\ldots, \mathbb{G}_p$ into $r$ successive subsequences
 $\scr{G}_1 = \{\mathbb{G}_1,\mathbb{G}_2,\ldots,\mathbb{G}_{rl}\},$ $\scr{G}_2 =\{\mathbb{G}_{rl+1},
 \ldots,\mathbb{G}_{2rl}\},$
 $\ldots$ $ \scr{G}_{r-1}=\{\mathbb{G}_{((r-2)rl +1},\ldots\mathbb{G}_{(r-1)rl}\},$
$\scr{G}_r =\{\mathbb{G}_{(r-1)rl+1},\ldots, \mathbb{G}_p\}$,
each of length   $r$  except for the last which must be of length  $p-l(r^2-r)\geq
lr$. Each of these $r$ sequences $\scr{G}_i,\;i\in \mathbf{r},$ consists of $r$ successive subsequences
which, in turn, are jointly strongly connected. Thus each of the $r$ composed graphs
$\mathbb{H}_1=\mathbb{G}_{rl}\circ\cdots\circ\mathbb{G}_1$,
 $\mathbb{H}_2= \mathbb{G}_{2rl}\circ\cdots\circ\mathbb{G}_{rl+1}$, $\ldots $,
  $\mathbb{H}_{r-1} = \mathbb{G}_{(r-1)rl}\circ\cdots\circ\mathbb{G}_{(r-2)rl+1}$,
 $\mathbb{H}_r = \mathbb{G}_p\circ\cdots\circ\mathbb{G}_{(r-1)rl+1}$ can be written
  as the composition of $r$ strongly connected graphs. But
the composition of any sequence
of $r$ or more strongly connected  graphs in $\scr{G}_{sa}$ is
 a complete graph \{cf. Proposition 4 of \cite{reachingp1}\}.
Thus each of the   graphs $\mathbb{H}_k,\;k\in\mathbf{r}$,
is a complete graph.
 Therefore each $\mathbb{H}_k$
 contains     every possible
  arc  $(i,j)$.  It follows that for any $i,j\in\mathbf{m}$ and any $k\leq r$,
   there must be a  route over the sequence $\scr{G}_k$ from $j$ to $i$.

Let $i_1,i_2,\ldots, i_{m}$ be any reordering of the sequence $1,2,\ldots, m$.  In the light of
 the discussion in the previous paragraph, it is clear that for each  $k\in\{1,2,\ldots r-1\}$,
  there must be a route 
$i_k = j_{(k-1)r },j_{ (k-1)r +1   },\ldots,j_{kr} = i_{k+1}$
over $\scr{G}_k$ from
 $i_k$ to $i_{k+1}$. Similarly there must be a route $i_r = j_{(r-1)r},j_{(r-1)r +1},\ldots,j_{q  } = i_{m}$
from $i_r$ to $i_m$ over $\scr{G}_r$.
Thus $i_1=j_1,j_2,\ldots, j_p=i_m$ must be route over  the
 overall sequence $\mathbb{G}_1,\mathbb{G}_2, \ldots, \mathbb{G}_p$. In view of Lemma \ref{gum},
 the matrix product
 $P_{j_p}\cdots P_{j_0}$ must be a component of the of the $i_mi_1$th block entry of
 $$M=P(S_p\otimes I)P(S_{p-1}\otimes I)\cdots  P(S_1\otimes I)P.$$
But $i_1,i_2,\ldots,i_m$ are distinct integers  and each appears in the sequence $j_0,j_1,\ldots,j_p$ at least once. Therefore the
$i_mi_1$th block entry of $M$ is complete.
Since this reasoning applies
for any sequence of $m$ distinct  vertex labels $i_1,i_2,\ldots, i_{m}$  from the set $\{1,2,\ldots, m\}$,
 every block entry
of $M$, except for the diagonal blocks, must be a complete projection matrix polynomial. If follows from Proposition \ref{ve}
and \rep{trouble2}
that $M$ is a contraction. $\qed $

\noindent{\bf Proof of Theorem \ref{nmeets}:} Let
$\scr{H}_i =Q_{(i-1)l+1},\ldots,Q_{il},\;i\in\{1,2,\ldots,\omega \}$,
 be any set of $\omega $ sequences in $\scr{C}_l$.
Since each $\scr{H}_i\in\scr{C}_{l}$,
each graph $\gamma(Q_{il}Q_{il-1}\cdots Q_{(i-1)l+1}),\;i\in\{1,2,\ldots,\omega \}$
is strongly connected. Therefore the sequence $\gamma(Q_1),\gamma(Q_2),\ldots, \gamma(Q_{\omega l})$ is
repeatedly $l$-connected.  Since there are $\omega l$ matrices in the $Q_i$ - sequence,
 Proposition \ref{ppo} applies.  Therefore  for any  set of sequences
  $\scr{H}_i\in\scr{C}_l,\;i\in\{1,2,\ldots,\omega\}$,
$||P(Q_{\omega l}\otimes I)P(Q_{\omega l-1}\otimes
 I)\cdots  P(Q_1\otimes I)P||<1$.  Since $\scr{C}_l$ is compact, $\lambda <1$.

  Set
 $M_t =P(S_t\otimes I)P(S_{t-1}\otimes I)\cdots  P(S_1\otimes I)P,\;\;\;t\geq 1$ and
$N_k = P(S_{\omega lk}\otimes I)P(S_{\omega l k-1}\otimes I)\cdots  P(S_{\omega l
 (k-1)+1}\otimes I)P,\;\;$
for  $k\in\mathbf{q}_t$,  where  $q_t$ is
 the unique integer quotient of $t$ divided by $\omega l$.
Since   $P^2=P$, it must be true that
$M_t =R_tN_{q_t}N_{q_t-1}\cdots N_1$
where  $R_t=P(S_t\otimes I)P(S_{t-1}\otimes I)\cdots  P(S_{q_tl +1}\otimes I)P$.
Since the sequences
$S_{ l(i-1)k +1},S_{ l(i-1)k +2},\ldots, S_{ lik},\;i\in\{1,2,\ldots, \omega\},$ $k\in\mathbf{q}_t,$
are all in $\scr{C}_l$,
 it must be true that  $||N_k||\leq\lambda^{\omega l},\;k\in\mathbf{q}_t$.
Therefore $||N_{q_t}N_{q_t-1}\cdots N_1||\leq\lambda^{\omega lq_t}$ so $||M_t||\leq ||R_t||\lambda^{\omega lq_t}$.
But for any $m\times m$ stochastic matrix $S$, $||S\otimes I|| = 1$ because $|S|_{\infty} = 1$.  In addition, $||P|| \leq 1$
because of \rep{snow1}. From  these observations and the fact that $||\cdot||$ is sub-multiplicative, it follows that
$||R_t||\leq 1$; thus
\eq{||M_t||\leq  \lambda^{\omega lq_t}.\label{op}}
Moreover $t = \omega lq_t +\rho_t$  where $\rho_t$ is the unique integer remainder of $t$ divided by $\omega l$.
Thus  $ \lambda^{\omega l q_t} = \lambda^{t-\rho_t}$.  But  $\rho_t < l\omega  $ and $\lambda <1$ so
$\lambda^{(t-\rho_t)} \leq \lambda^{(t-l\omega)}$.
It follows from this and \rep{op}  that \rep{limits} is true. $\qed $

\subsection{Convergence Rate}\label{CR} In this section we will justify the claim that the expression for $\lambda$ given by
\rep{rate}
 is a worst case bound on the \{geometric\} convergence rate  for the algorithm \rep{a1} for the case when $Ax=b$ has a unique solution and all of the neighbor graphs  encountered are strongly connected. To establish this claim we will need a lower bound on the coefficients  of the nonzero $n\times n$  projection matrix polynomials which comprise the
  $ m\times m$ partition of
 $P(F_q\otimes I)P(F_{q-1}\otimes I)\cdots P(F_1\otimes I)P$ . The bound is given  next.
  \vspace{.1in}

\begin{lemma}Let $s$ be a positive integer and suppose that the nonzero block projection matrix
$$M_{ij}=\sum_{k=1}^d \lambda_k P_{h_k(1)}P_{h_k(2)}\cdots P_{h_k(s+1)}$$
is the $ij$th submatrix      within the
 $nm\times nm$ matrix
$M=P(F_s\otimes I)P(F_{s-1}\otimes I)\cdots P(F_1\otimes I)P$
where $d$ is a positive integer, each $h_k(i)$ is an integer
in $\mathbf{m}$  and each $\lambda_k$ is a positive number.
Then
$$\lambda_k \geq \frac{1}{m^s},\;\;\;k\in\mathbf{d}.$$
\label{bbx}\end{lemma}

\noindent
{\bf Proof of Lemma \ref{bbx}:}
We will prove the lemma by induction on $s$.
Suppose first that $s=1$. Then $M=P(F_1\otimes I)P$ and $M_{ij} = f_{ij}P_iP_j$ where $f_{ij}$ is the $ij$th entry in $F_1$. Since $M_{ij}\neq 0$, $f_{ij} \neq 0$.
Since $F_1$ is a flocking matrix,
each nonzero entry  is  bounded below by $\frac{1}{m}$.
Thus, in this case the lemma is clearly true.

Now suppose that the lemma holds for all $s$ in the range $1\leq s \leq p$ where $p\geq 1$ is an integer.
 Let $s=p+1$. Then $M=P(F_s\otimes I)N$ where
$N=P(F_
{s-1}\otimes I)P(F_{s-2}\otimes I)\cdots P(F_1\otimes I)P$.
Thus, for all $i,j\in\mathbf{m}$,
\eq{M_{ij} = \sum_{k=1}^m f_{ik}P_iN_{kj}\label{prod}}
where $f_{ik}$ is the $ik$th entry of $F_q$ and $N_{kj}$ is the $kj$th block entry of $N$.
Each  $N_{kj}$ is either the $n\times n$ zero matrix
or a projection matrix polynomial of the form
$$N_{kj}=\sum_{l=1}^c \lambda_l P_{h_l(1)}P_{h_l(2)}\cdots P_{h_l(p+1)}$$
where $c$ is a positive integer, each $h_l(i)$ is an integer
in $\mathbf{m}$, and for all $l\in\mathbf{c}$,
$\lambda_l > 0$.  Thus if $N_{kj}\neq 0$, then $\lambda_l \geq  \frac{1}{m^p}$ because of the inductive hypothesis.
From \rep{prod},
$$M_{ij} = \sum_{k=1}^m \sum_{l=1}^c  \left(f_{ik}\lambda_l \right) P_iP_{h_l(1)}P_{h_l(2)}\cdots P_{h_l(p+1)}.$$
Since $F_s$ is a flocking matrix,
either $f_{ik}=0$ or $f_{ik}\ge\frac{1}{m}$,
which implies that either $f_{ik}\lambda_l=0$ or $f_{ik}\lambda_l\ge \frac{1}{m^{p+1}}$. Since $M_{ij}\neq 0$, it must therefore be a projection matrix polynomial  whose   coefficients
 are all bounded below by $\frac{1}{m^{p+1}}$.
Thus the lemma holds for $s=p+1$. By induction, the lemma
is established and the proof is complete.
$\qed$

\noindent{\bf Proof of Corollary \ref{4th}:} To prove this corollary, it is sufficient to show  that for any set of $q$ flocking matrices $F_1,F_2,\ldots,F_q$, the mixed matrix norm of the matrix
 $$M = P(F_q\otimes I)P(F_{q-1}\otimes  I)\cdots P(F_1\otimes I)P$$
 satisfies
 \eq{||M|| \leq 1-\frac{(m-1)(1-\rho)}{m^q}\label{frt}}
where  $\rho$ is given by \rep{rate2}.
 By definition
\eq{||M|| = \max_{i\in\mathbf{m}}\left (\sum_{j=1}^m|M_{ij}|_2\right )\label{norman}}
where $M_{ij}$ is the $ij$th block entry of $M$.
In view of \rep{trouble2}, the nominal bound of $M$ is the stochastic matrix
 $ F_qF_{q-1}\cdots F_1$.  Thus
 \eq{|M_{ij}|_2\leq f_{ij}\label{dop1}}
   where $f_{ij}$ is the $ij$th entry in $F_qF_{q-1}\cdots F_1$.

Fix $i,j\in\mathbf{m}$ with $i\neq j$. As noted just at the end of the proof of Proposition \ref{ppo},
each  block entry of $M$, except for the diagonal blocks, must be a complete projection matrix polynomial.
Thus
 $M_{ij}$  must be  a nonzero matrix of form
$$ M_{ij} =\sum_{k=1}^{d}\lambda_k P_{h_k(1)}P_{h_k(2)}\cdots P_{h_k(q+1)}$$
where  $d$ is a  positive integer, each $\lambda_k $ is a real positive
   number, and  each
 $h_k(i)$  is an integer in $\mathbf{m}$. Completeness   also means that for some integer $s\in\mathbf{d}$,
 each of the matrices in the set $\{P_1,P_2,\ldots, P_m\}$ appears in the product
 $P_{h_s(1)}P_{h_s(2)}\cdots P_{h_s(q+1)}$ at least once; consequently \newline $P_{h_s(1)}P_{h_s(2)}\cdots P_{h_s(q+1)}\in\scr{C}$ so $|P_{h_s(1)}P_{h_s(2)}\cdots P_{h_s(q+1)}|_2\leq \rho$.
  In addition,
  $|P_{h_k(1)}P_{h_k(2)}\cdots P_{h_k(q+1)}|_2\leq 1,\;k\in\mathbf{d}$
  because of  Lemma \ref{crack}.
   It follows that
 \begin{eqnarray*}|M_{ij}|_2 & \leq &\sum_{k=1}^{d}\lambda_k |P_{h_k(1)}P_{h_k(2)}\cdots P_{h_k(q+1)}|_2\\
   &= &
\sum_{\matrix{k=1, k\neq s }}\lambda_k |P_{h_k(1)}P_{h_k(2)}\cdots P_{h_k(q+1)}|_2 +\lambda_{s} |P_{h_s(1)}P_{h_s(2)}\cdots P_{h_s(q+1)}|_2 \\
&\leq &
\sum_{\matrix{k=1, k\neq s }}\lambda_k +\lambda_{s}\rho  \\
&=&\sum_{k=1 }^d\lambda_k -\lambda_s(1 -\rho).  \\
\end{eqnarray*}
Recall that $\sum_{k=1}^d\lambda_k$ is the nominal
  bound of $M_{ij}$; thus
   $\sum_{k=1}^d\lambda_k = f_{ij}$. Meanwhile, by Lemma \ref{bbx}, $\lambda_s\geq \frac{1}{m^q}$. If follows that
   \eq{|M_{ij}|_2\leq f_{ij}-\frac{1}{m^q}(1- \rho).\label{dups}}

Now for each $i\in\mathbf{m}$,
$$\sum_{j=1}^m|M_{ij}|_2 = \sum_{\matrix{j=1, j\neq i}}^m|M_{ij}|_2 +|M_{ii}|_2.$$
From \rep{dop1} and \rep{dups} it follows that
$$\sum_{j=1}^m|M_{ij}|_2 \leq  \sum_{\matrix{j=1, j\neq i}}^m \left (f_{ij} - \frac{1}{m^q}(1- \rho)\right )  +f_{ii}.$$
Clearly
$$\sum_{j=1}^m|M_{ij}|_2 \leq  1 - \frac{(m-1)}{m^q}(1- \rho). $$
From this and \rep{norman} it follows that \rep{frt} is true. $\qed $

\section{Tracking}\label{tracking}

An especially important consequence of exponential  convergence is that it enables a slightly modified version of algorithm \rep{a1} to track
the solution to $Ax = b$  with ``small error'' when $A$ and $b$ are changing  with time, provided the rates at which  $A$ and $b$  change are  sufficiently small. In the sequel we sketch why this is so for the case when the time-varying equation $A(t)x(t) = b(t)$ has a unique solution for every fixed value of  $t$.
We continue to follow the agreement principle stated at the beginning of \S \ref{alg}. In particular,
suppose that at each time $t$   agent $i$ knows the pair $(A_i(t+1),b_i(t+1))$  and using it, computes any solution $z_i(t)$ to $A_i(t+1)x = b_i(t+1)$  such as $A_i(t+1)'(A_i(t+1)A_i'(t+1))^{-1}b_i(t+1)$, if
 $A_i(t+1)$ has linearly independent rows.
If
$K_i(t)$ is a basis matrix for the kernel of $A_{i}(t+1)$ and  we restrict the updating  of $x_i(t)$ to iterations of the form
$x_i(t+1) = z_i(t) + K_i(t)u_i(t),\;\;t\geq 1$,
then no matter what $u_i(t)$ is,    each $x_i(t+1)$ will   satisfy $A_i(t+1)x_i(t+1) = b_i(t+1),\;t\geq 1$.
Just as before, and for the same reason, we will choose  $u_i(t)$ to minimize the difference
 $\left (z_i(t)+K_i(t)u_i(t)\right ) - \frac{1}{m_i(t)}\left(\sum_{j\in\scr{N}_i(t)}x_j(t)\right )$ in the least squares sense.
Doing this leads at once to an iteration for
 agent $i$
  of the form
\eq{x_i(t+1) = z_i(t)- \frac{1}{m_i(t)}P_i(t)\left (  m_i(t)z_i(t)-\sum_{j\in\scr{N}_i(t)}x_j(t)\right ),t\geq 1\;\;\;\label{newa1}}
where for each $t\geq 0$, $P_i(t)$ is the time-varying  orthogonal projection on the kernel of $A_i(t+1)$ and $x_i(1)$ is a solution to $A_i(1)x = b_i(1)$. It is worth noting that even though $z_i(t)$  is not uniquely specified here, update rule
\rep{newa1} is because $(I-P_i(t))z_i(t)$ is independent of the choice of $z_i(t)$, just as it was in the time-invariant case discussed earlier.
 The algorithm just described, differs from \rep{a1} in two respects.  First the $P_i$ are now time dependent and second, instead of using $x_i(t)$ to represent a preliminary  estimate of the solution to   $A(t+1)x = b_i(t+1)$, we use $z_i(t)$ instead. This modification has the advantage of yielding  an algorithm which is much easier to analyze
 than would be the case were we  to use $x_i(t)$.

  We will assume that $A(t)$ and $b(t)$ are uniformly bounded signals and   for simplicity, we will further assume that each $A_i(t)$ has full row rank for all $t$; more specifically we will require the determinant of
  $A_i(t)A_i'(t)$ to be bounded away from $0$ uniformly. We will also assume that  $A(t+1) = A(t)+ \delta_A(t),\;t\geq 1$ and $b(t+1) = b(t)+\delta _b(t),\;t\geq 1$  where $\delta_A(t)$ and $\delta _b(t)$
are small norm  bounded signals. Since
  $P_i(t) = I-A_i'(t+1)(A_i(t+1)A_i'(t+1))^{-1}A_i(t+1)$, $P_i(t)$ will be uniformly bounded.  Note that it is possible to write
  $P_i(t+1) = P_i(t) + \bar{E}_i(\delta_A(t+1)),\;t\geq 0$ where $\bar{E}_i(\cdot) $ is a continuous function  satisfying $\bar{E}_i(0) =0$.

Our goal is to explain why this algorithm can track  the unique solutions $x^*(t)$ to  $A(t)x(t) = b(t)$.
  As a first step, observe that
$x^*(t+1) = x^*(t) - \delta(t)$  where $\delta(t) = \delta_A(t)A^{-1}(t)b(t)-A^{-1}(t+1)\delta_b(t)$.
Clearly
\eq{x^*(t+1) = x^*(t)- \frac{1}{m_i(t)}P_i(t)\left (  m_i(t)x^*(t)-\sum_{j\in\scr{N}_i(t)}x^*(t)\right )-\delta(t)\label{a1n}}
for $t\geq 1$ because the term in parentheses on the right of this equation  is zero.
 Thus if we define the error signal
 \eq{e_i(t) = x_i(t)-x^*(t),\;\;\;i\in\mathbf{m},\;\;t\geq 1\label{redefinenn}}
 then $P_i(0)e_i(1) = e_i(1)$ and
\begin{eqnarray*}
e_i(t+1)& =& (I-P_i(t))(z_i(t)-x^*(t+1))+\frac{1}{m_i(t)}P_i(t)\sum_{j\in\scr{N}_i(t)}e_j(t)+P_i(t)\delta(t) ,\;\;t\geq 1.
\end{eqnarray*}
 But since both $x^*(t+1)$ and $ z_i(t)$ are solutions to $A_i(t+1)x = b_i(t+1)$, the vector $z_i(t)-x^*(t+1)$ is in the kernel
 of $A_i(t+1)$; this  implies that $(I-P_i(t))(z_i(t)-x^*(t+1)) = 0$.  It follows that

$$e_i(t+1) = \frac{1}{m_i(t)}P_i(t)\sum_{j\in\scr{N}_i(t)}e_j(t) +P_i(t)\delta(t)\;\;t\geq 1,\;\;i\in\mathbf{m}.$$
Hence if we again define
 $e(t) = {\rm column}\{e_1(t),e_2(t),\ldots, e_m(t)\}$ there results
\eq{e(t+1) =P(t)(F(t)\otimes I)e(t) +P(t) (\mathbf{1}\otimes \delta(t)),\;\;t\geq 1\label{pupnew}}
where for $t\geq 0$,
 $P(t)$ is the $mn\times mn$ matrix $P(t) = {\rm diagonal}\{P_1(t),P_2(t),\ldots, P_m(t)\}$, $\mathbf{1}$ is the  $m$ vector of $\mathbf{1}$'s, and
and for $t\geq 1$, $F(t)$ is  the same flocking matrix used earlier. Observe that since $P^2(t) = P(t)$, \rep{pupnew} implies that
 $P(t)e(t+1) = e(t+1),\;t\geq 1$; thus  $P(t-1)e(t) = e(t),\;t\geq 2$. But  $P(0)e(1) =e(1)$ because $P_i(0)e_i(1) = e_i(1)$ as was noted earlier. Therefore $P(t-1)e(t) = e(t),\;t\geq 1$.
 If we define   $E(t) = {\rm diagonal}\{\bar{E}_1(\delta_A(t)),\bar{E}_2(\delta_A(t)),\ldots, \bar{E}_m(\delta_A(t))\},\;t\geq 1$,
then  $E(t)$ will have a small norm if $\delta_A(t)$ does. In view of the definition of $E(t)$,  $P(t) = P(t-1) +E(t),\;t\geq 1$.
Clearly for $t\geq 1$, $P(t)e(t) = P(t-1)e(t) +E(t)e(t)$ so $P(t)e(t) = e(t) +E(t)e(t)$.
 Therefore
\begin{eqnarray}e(t+1) &=&(P(t)(F(t)\otimes I)P(t)-  P(t)(F(t)\otimes I)E(t))e(t)+P(t)( \mathbf{1}\otimes \delta(t)),\;t\geq 1.\label{pupnewer}\end{eqnarray}

We claim that for $|\delta_A(t)|_{2}$ sufficiently small for all $t$, the time varying matrix
$$P(t)(F(t)\otimes I)P(t)-P(t)(F(t)\otimes I)E(t)$$
is exponentially stable  assuming the sequence of neighbor graphs $\mathbb{N}(t),\;t\geq 1$  satisfies the hypotheses  of Theorem \ref{mainer}.
Because $|P(t)(F(t)\otimes I)E(t)|_2$  will be small if $|\delta_A(t)|_{2}$ is, to establish exponential stability, it is sufficient to show that the matrix $P(t)(F(t)\otimes I)P(t)$ is exponentially stable for  $|\delta_A(t)|_{2}$ sufficiently small.
To do this it is convenient to first consider the matrix
 $M(t,s)=P(s)(F(t)\otimes I)P(s)$.
  We know already
 that for every fixed value of $s$, the linear system $z(t+1)= M(t,s)z(t)$ has a unique equilibrium  at the origin.
 In view of Theorem \ref{nmeets} we also know that every solution to this equation tends to the origin exponentially fast.  In other words, for each fixed $s$, $M(t,s)$ is an exponentially stable time varying matrix. Our goal is to show that
 $M(t,t)$ is exponentially stable as well provided  $|\delta_A|_2$  is sufficiently small. While doing  this  is actually a fairly straightforward exercise in linear system theory, it is nonetheless a little bit unusual and so for the sake of clarity we will proceed.

The key fact we will use, which comes  from basic Lyapunov theory, is that for every constant $nm\times nm$ matrix $B$ and every fixed value of $s$, the matrix
$$L(t,s,B)=\sum_{\tau=t}^{\infty}\Phi_s'(\tau,t)B\Phi_s(\tau,t)$$
is a uniformly bounded function of $t$, where $\Phi_s(t,\tau)$  is the state transition matrix of $M(t,s)$.
This is an immediate consequence of exponential stability. It is also true, and is easily verified,  that $L(t,s,B)$ satisfies the Lyapunov
equation
\eq{L(t,s,B) = M'(t,s)L(t+1,s,B)M(t,s) +B,\;\;t\geq 1\label{ly}}
 for all $s\geq 0$. We use these observations in the following way.

Let $Q(t,s) = L(t,s,I)$.  Then by a straightforward but tedious computation using \rep{ly},
$$Q(t,s+1) - Q(t,s) = \Delta_Q(t,s,\delta_A(s))$$
where
$\Delta_Q(t,s,\delta_A)$ is a bounded  function of $t$ and $s$ and a continuous function of $\delta_A$
satisfying $\Delta_Q(t,s,0) = 0,\;t,s\geq 0.$
Observe that
\begin{eqnarray*}
  Q(t,s) &=& M'(t,s)(Q(t+1,s+1)M(t,s) +I\\
  &&- M'(t,s)\Delta_Q(t,s,\delta_A(s))M(t,s).
\end{eqnarray*}
Thus if the uniform norm bound on  $|\delta_A(t)|_{2}$ is small enough, then
 $I- M'(t,t)\Delta_Q(t,t,\delta_A(t))M(t,t)$ will be positive definite
implying that
$$Q(t,t)- M'(t,t)Q(t+1,t+1)M(t,t) $$ is negative definite for all $t$ and thus that $z'Q(t,t)z$ is a valid Lyapunov
 function for the equation $z(t+1) = M(t,t)z(t)$.
 Therefore the time varying matrix
$P(t)(F(t)\otimes I)P(t)-P(t)(F(t)\otimes I)E(t)$ will be an exponentially stable matrix if the norm bound on $\delta_A(t)$ is sufficiently small.

Of course $\delta $ will be small in norms if both $\delta_A$ and $\delta_b$ are.  From this and the exponential stability of the system  \rep{pupnewer}, it follows that for sufficiently slow variations in $A$ and $b$, $e$ will be small and in this sense,
each of the $x_i(t)$ will eventually track  with small error, the  time-varying solution $x^*(t)$ to $A(t)x^*(t) = b(t)$. Exponential stability is the key property upon which this conclusion rests.

These observations prompt one to ask a number of questions:  How small must  $\delta_A$ be for tracking to occur and
what is the ``gain'' between the sum of the  norms of $\delta_A$ and $\delta_b$ and the norm of the tracking error
$e$?  In the event that $\delta_A$ and $\delta_b$ can be regarded as solutions to neutrally stable linear recursion equations, can the internal model principal \cite{imp} be used to modify the algorithm so as to achieve a zero tracking error asymptotically? There are questions for future research. \vspace{.1in}

\noindent{\bf Example:} The following example  is intended to  illustrate the tracking capability of the algorithm just discussed. The  equation to be solved is $A(t)x(t) = b$ where for $t\geq 1$
$$A(t) = \matt{2 & 3& 5\cr 4 & 9 & -8\cr 1 &5 & 10}+\sin 0.1(t-1)\matt{.1 &.09 &-.24\cr .2 &-.6 & .1\cr .03& .05 & .4}$$ and $$b = \matt{10 \cr 5\cr 16}+ \sin 0.6(t-1)\matt{.1 \cr .2 \cr .3}.$$
Agent $i$ knows the $i$th row of the matrix $\matt{A(t) & b(t)}$  at time $t-1$ and initializes its state $x_i(t)$ as follows.
$$x_1(1) = \matt{11.5\cr -1\cr -2}\hspace{.3in}x_2(1) = \matt{ 1.25\cr 0\cr 0},\hspace{.3in} x_3(1) = \matt{-9\cr 1\cr 2}$$
and  $z_i(t-1) =A_i'(t)(A_i(t)A_i'(t))^{-1}b_i(t),\;i\in\mathbf{3}$.
A plot of the evolution of the two norm  of the
tracking error $e(t)$ is shown in the following figure.

\begin{figure}[h]\centerline{
 \includegraphics[height = 2.2in]{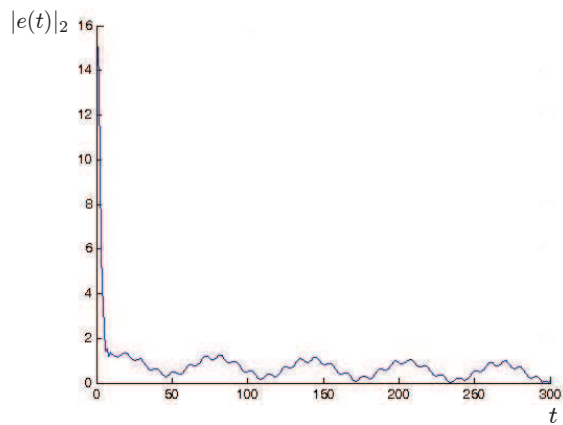}}
    \label{bbb6r}
    \caption{$|e(t)|_2$ vs $t$}
   \end{figure}

\section{Asynchronous Operation}\label{asynch}

 In this section we show that with minor modification, the algorithm we have been studying, namely \rep{a1}, can be implemented
asynchronously. The relevant update rules are given by \rep{kop}.  Since these rules are defined with respect to different and unsynchronized time  sequences, for convergence analysis one needs to derive  a  model on which all update rules
evolve synchronously  with respect to a single time scale.   Such a model is given by \rep{nna2r}. Having accomplished this, we then establish the  correctness of \rep{kop}, but only for the case
  when there are no communication delays.
 The more realistic version of the problem in which delays are explicitly taken into account is treated in \cite{asylineareqn}. The ideas exploited there closely parallel those used to analyze the asynchronous
 version of the unconstrained consensus problem treated in \cite{async}.

Let $t$ now take values in the real time interval $[0.\infty)$. We begin by
   associating with each  agent $i$,
 a strictly increasing,  infinite sequence  of {\em event times} $t_{i1}, t_{i2},\ldots $  with the understanding that $t_{i1}$ is the time agent $i$ initializes its state and the remaining $t_{ik},\;k>1$ are the times at which agent $i$ updates its state. Between any two successive event times  $t_{ik}$ and $t_{i(k+1)}$,
  $x_i(t)$ is held constant.
   We assume that  for any $k\geq 1$, $x_i(t)$ equals its limit from above as $t$  approaches
  $t_{ik}$; thus  $x_i(t)$ is constant on each open half interval
  $[t_{ik},t_{i(k+1)}),\;\;k\geq 1$.

  We assume that  for $i\in\{1,2,\ldots,m\}$,  agent  $i$'s event times  satisfies \eq{\bar{T}_i \geq  t_{i(k+1)} -t_{ik} \geq T_i,\;\;\;\;k\in\{1,2,\ldots\}\label{p1} }
   where $\bar{T}_i$ and $T_i$ are  positive numbers such that  $\bar{T}_i>T_i$.
  Thus the event times of agent $i$ are distinct and the difference between any two successive event times cannot be too large. We make no assumptions at all  about the relationships between the event times of different agents. In particular,  two agents may have completely different unsynchronized event time sequences.

We assume, somewhat unrealistically, that at each of its event times  $t_{ik}$, agent $i$ is able to acquire the state $x_j(t_{ik})$ of each of its ``neighbors'' where by a {\em neighbor } of agent $i$ at time $t_{ik}$ is  meant
any agent in the network whose state is available to agent $i$  at time $t_{ik}$.  In the more realistic version of the problem treated in \cite{asylineareqn}, it is assumed that $x_j(t_{ik})$ is only available to agent $i$ after a delay
which accounts both for  transmission time and the fact that the time at which $x_j(t_{ik})$ is acquired
is typically some time  in between  $t_{ik}$ and one of agent $i$'s subsequent event times. There are some subtle issues here in setting up an appropriate  model;
we refer the reader to \cite{asylineareqn} for an explanation of what they are and how they are addressed.

In the sequel, for $k>1$  we write  $\scr{N}_{i}(t_{ik})$ for
      the set of labels of agent $i$'s neighbors at time $t_{ik}$ while $k = 1$  we define   $\scr{N}_{i}(t_{i1}) = i$.  Since agent $i$ is always taken to be a neighbor of itself,
   $\scr{N}_{i}(t_{ik})$ is never empty.

Prompted by \rep{a1},
  the  update rule for agent $i$ we want to consider for the asynchronous case  is
 \eq{x_i(t_{i(k+1)}) = x_i(t_{ik})- \frac{1}{m_i(t_{ik})}P_i\left (  m_i(t_{ik})x_i(t_{ik})-
\sum_{j\in\scr{N}_i(t_{ik})}x_j(t_{ik})\right )\label{kop}}
where $k\geq 1$, and for $j\in\scr{N}_i(t_{ik})$,  $m_i(t_{ik})$ is the number  of labels in
$\scr{N}_{i}(t_{ik})$, and as before,
$P_i$ is the orthogonal projection  on the kernel of $A_i$.

 To proceed we need  a common time scale on which all $m$ agent update rules can be defined.  For this,
  let  $t_1 = \max_i\{t_{i1}\}$   and write
   $\scr{T}_i$ for  the event times  of agent $i$ which are greater than or equal to $t_1$.
 Let    $\scr{T}$ denote  the  set of all event times  of all  $m$ agents which are greater
  than or equal to $t_1$. Thus $\scr{T}$
   is the union of the $\scr{T}_i$.  Relabel the times in $\scr{T} $ as $t_1,t_2,\ldots, t_p,\ldots $
    so that $t_p<t_{p+1}$ for $p\geq 1$.
We define the {\em extended neighbor set}
  of agent $i$, written
 $ \bar{\scr{N}}_i(p)$, to be $\scr{N}_i(t_p)$  if $t_p$ is an event time of agent $i$.
 For times  $t_p\in\scr{T}$  which are not event times of agent $i$, we
 define $\bar{\scr{N}}_i(p) = \{i\}$.  Doing this enables us to extend the domain of applicability
 of update rule \rep{kop}  from $\scr{T}_i$ to all of $\scr{T}$. In particular, for $p\geq 1$,
\eq{x_i(t_{p+1}) = x_i(t_{p})- \frac{1}{\bar{m}_i(p)}P_i\left (  \bar{m}_i(p)x_i(t_{p})-
\sum_{j\in\bar{\scr{N}}_i(p)}x_j(t_{p})\right )\label{nna2r}}
where $\bar{m}_i(p)$ is the number of indices in $\bar{\scr{N}}_i(p)$.
The validity of this formula is a simple consequence of the assumption that for $i\in\{1,2,\ldots, m\}$,  $x_i(t)$ is constant on each open half interval
  $[t_{ik},t_{i(k+1)}),\;\;k\geq 1$.

Observe that  \rep{nna2r} is essentially the same as update rule \rep{a1} except that extended neighbor sets
replace the original neighbor sets.
As with the synchronous case, convergence depends on connectivity of  the graphs  determined
by the neighbor sets upon which update rules \rep{nna2r} depend. Accordingly,
 for each $p\geq 1$
   we  define the {\em extended neighbor graph} $\bar{\mathbb{N}}(p)$
    to be that directed graph in $\scr{G}_{sa}$ which has an arc
   from vertex $j$ to vertex $i$ if $j\in\bar{\scr{N}}_i(p)$.
The following  is an immediate consequence of Theorem \ref{mainer}.
\begin{theorem} Suppose each agent $i$ updates its state $x_i(t)$ according to  rule \rep{kop}.
Suppose in addition that for some positive integer $l$, the sequence of extended neighbor graphs
 $\bar{\mathbb{N}}_i(p),\; p\geq 1$ is repeatedly jointly strongly connected. Then
 there exists a positive constant
$\lambda<1$ for which
 all $x_i(t_p)$ converge to the same solution to $Ax=b$ as $p\rightarrow\infty$, as fast as $\lambda^p$ converges to $0$.
\label{amainer}\end{theorem}
Perhaps of greatest interest is the situation when  the original neighbor graph $\mathbb{N}(t)$ is
 independent of time. In this case it is possible to address convergence without reference to
  extended neighbor graphs.

\begin{corollary} Suppose that the original neighbor graph  $\mathbb{N}(t)$ is
 independent of time and strongly connected.
Suppose each agent $i$ updates its state $x_i(t)$ according to  rule \rep{kop}.  Then
 there exists a positive constant
$\lambda <1$ for which
 all $x_i(t_p)$ converge to the same solution to
  $Ax=b$ as $p\rightarrow\infty$, as fast as $\lambda^p$ converges to $0$.\label{bain}\end{corollary}
  The proof of Corollary \ref{bain} depends on the following lemma.

\begin{lemma} Suppose that the original neighbor graph $\mathbb{N}(t)$ is a constant graph $\mathbb{N}$.
For $i\in\mathbf{m}$, let $\bar{T}_i$ be an upper bound on the difference between
 each pair of  successive event times of agent $i$.
 Then for any pair of event times $t_a,t_b\in\scr{T}$ satisfying
 $t_b-t_a\geq \max\{\bar{T}_1,\bar{T}_2,\ldots ,\bar{T}_m\}$,
 $\mathbb{N}$ is a spanning subgraph of the composed graph
  $\bar{\mathbb{N}}(b)\circ\bar{\mathbb{N}}(b-1)\cdots
 \circ\bar{\mathbb{N}}(a)$.\label{last}\end{lemma}

\noindent{\bf Proof of Lemma \ref{last}:} Let $\scr{N}_i$ denote the neighbor set of agent $i$.
For $i\in\mathbf{m}$,  $t_{i(j+1)}-t_{ij}\leq\bar{T}_i\leq t_b-t_a,\; j\geq 1$. Therefore  the set $\{t_{a},t_{(a+1)},\ldots, t_{b}\}$
must contain at least one event time $t_{p_i}$ of each agent $i$. Since $\bar{\scr{N}}_i(p_i) =
\scr{N}_i,\;i\in\mathbf{m}$, for each $j\in\scr{N}_i$ there must be an arc from $j$ to $i$ in
 $\bar{\mathbb{N}}(p_i)$. It follows  from the definition of $\mathbb{N}$, that its arc set
 must be contained in the union of the arc sets of the graphs  $\bar{\mathbb{N}}(a),\bar{\mathbb{N}}(a+1),
 \ldots, \bar{\mathbb{N}}(b)$.
But the arc set of the union  of a finite number of graphs in $\scr{G}_{sa}$ is always a subset
 of the arc set of their composition \cite{reachingp1}.  Therefore the lemma is true. $\qed $


\noindent{\bf Proof of Corollary \ref{bain}:} Set $T_{\max} = \max\{\bar{T}_1,\bar{T}_2,\ldots ,\bar{T}_m\}$
and  $T_{\min} =\min \{T_1,T_2,\ldots, T_m\}$ and let
 $q$ be any positive integer for which $T_{\max}\leq qT_{\min}$.
Let $a$ and $b$ be positive integers satisfying
 $b-a = mq$.  We claim that $t_b-t_a\geq T_{\max}$.  To prove that this is so,
 suppose the contrary, namely that $t_b-t_a <T_{\max}$.  Then $t_b-t_a < qT_{\min}$.  But for each
  $i\in\mathbf{m}$, $T_{\min}$ is no larger than the time between
   any two successive event times of agent $i$. Thus  the closed interval $[t_a,t_b]$  must contain at most
    $q$
   event times of agent $i$. Since there are $m$ agents, $[t_a,t_b]$ must contain at most $mq$ event times.
   Therefore $b-a < mq  $ which is a contradiction.

In view of the preceding, $t_{b} - t_{a}\geq T_{\max}$ for any  positive integers $a$ and $b$ satisfying $b-a= mq$.
Therefore, by Lemma \ref{last}, $\mathbb{N}$ must be a spanning subgraph of the composed graphs
 $\bar{\mathbb{N}}(b)\circ\bar{\mathbb{N}}(t_{b-1})\cdots
 \circ\bar{\mathbb{N}}(a)$ for all such $a$ and $b$.  But $\mathbb{N}$ is strongly connected so each such composed
 graph   must be strongly connected as well.  Therefore the sequence of graphs
 $\bar{\mathbb{N}}(1),\bar{\mathbb{N}}(2),\ldots $ is repeatedly jointly strongly connected by successive
 subsequences of length $mq$.
 From this and  Theorem \ref{amainer} it follows that Corollary \ref{bain} is true. $\qed $

\section{Least Squares}\label{LS}

A limitation of the algorithm we have been discussing is that it is only applicable to linear equations for which there are solutions. In this section we explain how to modify the algorithm so that it can obtain least squares solutions to $Ax = b$ even in the case when $Ax = b$ does not have a  solution.  As before, we will approach the problem using standard consensus concepts rather than the more restrictive  concepts  based on distributed averaging.
To keep things simple, we will assume that the $A_i$ are full column rank matrices.

By the {\em least squares solution} to $Ax = b$ is meant a value of $x$ for which $A'Ax = A'b$. As is well known,
least squares solutions always exist, even if $Ax = b$ does not have a solution.
   It is very easy to verify that  a common least squares  solution $x$  to all
   of the agent equations $A_jx = b_j,\;j\in\mathbf{m}$ will not exist unless
 $Ax=b$  has a solution. Thus if a decentralized  least squares  solution  to $Ax = b$ is to be obtained in accordance with the agreement principle,  then each agent   must solve a different problem. To understand what that problem might be, consider for example the situation in which there are three agents. Suppose that
 the state $x_i$ of agent $i$ is augmented with two additional  $n$-vectors, namely $y_i$ and $z_i$ and that
 agents  $1$,  $2$  and $3$  are tasked  to solve the linear equations
  \begin{eqnarray*}A_1'A_1x_1 +y_1 &=& A_1'b_1\label{po1}\\
A_2'A_2x_2 +z_2 &=& A_2'b_2\label{po2}\\ A_3'Ax_3 -y_3 - z_3 &=& A_3'b_3\label{po3}\end{eqnarray*}
 respectively.  As we will show, it is
always possible for the agents to do this and the same time  to obtain values of the $x_i, y_i$ and $z_i$ for which
the three augmented state vectors $\bar{x}_i=\matt{x_i'&y_i'&z_i'}',\;i\in\mathbf{3}$ are the same.

 The existence of  a vector  $\bar{x}= \matt{x'&y'&z'}'$ for which  $\bar{x}_i= \bar{x},\;i\in\mathbf{3},$
 is  equivalent to the existence to a solution to the
equations $A_1'A_1x +y =A_1'b_1$,
$A_2'Ax +z =A_2'b_2$, and  $A_3'A_3x -y - z =A_3'b_3$.
In matrix terms, existence amounts to asking whether or not the equation
$M\bar{x} = q$
has a solution where  
 $$M = \matt{A_1'A_1 & I & 0 \cr A_2'A_2 & 0 & I\cr A_3'A_3 & -I & -I}\;\;\;\;{\rm and}\;\;\;\;\;\;q = \matt{A_1'b_1   \cr A_2'b_2 \cr A_3'b_3}.$$
Note that by simply adding block rows  block rows $1$ and $2$ of $\matt{ M & q}$  to block row $3$, one obtains
the matrix $\matt{\bar{M} & \bar{q}}$ where
$$\bar{M}=\matt{A_1'A_1 & I & 0 \cr A_2'A_2 & 0 & I\cr A_1'A_1+ A_2'A_2 +  A_3'A_3 & 0 & 0} $$ and $$\bar{q} =  \matt{A_1'b_1   \cr A_2'b_2 \cr A'_1b_1 +A'_2b_2 + A_3'b_3}.$$
Clearly the set of solutions to $M\bar{x} = q$ is the same as the set of solutions to $\bar{M}\bar{x} = \bar{q}$
because the matrices  $\matt{M & q}$  and $\matt{\bar{M} & \bar{q}}$ are row equivalent.   It is obvious that $\bar{M}$ has linearly independent columns because $A_1'A_1+ A_2'A_2 +  A_3'A_3$ is nonsingular; therefore $\bar{M}$ is nonsingular. As a result, a solution to $M\bar{x} = q$ must exist. Note in addition, that since such a solution must
 also satisfy $\bar{M}\bar{x} = \bar{q}$, $x$ must satisfy $(A_1'A_1+ A_2'A_2 +  A_3'A_3)x = A_1'b_1 +A_2'b_2+A_3'b_3$ which is the  least squares equation $A'Ax = A'b$. Therefore $x$ solves the least squares problem.

Recall that the idea exploited earlier in the paper for  crafting an algorithm for solving $Ax=b$,  was that if each agent $i$  were able to compute a solution $x_i$  to its own equation $A_ix_i = b_i$ and at the same time all agents were able to reach a consensus in that all $x_i$ were equal, then automatically each $x_i$ would necessarily satisfy $Ax_i = b$. This led at once to
 the linear iterations \rep{a1} which provide distributed solutions to $Ax = b$.  Since with obvious modification, the same idea
 applies to the least squares problem under consideration here, it is clear that the same approach will lead to
 linear iterations which provide a distributed solution to the least squares equation $A'Ax = A'b$. The update equations in this case are identical with those in \rep{a1} except that in place of  and $x_i$ and $P_i$ one would use the
 $\bar{x}_i$ and  $\bar{P}_i$  where $\bar{P}_i$  is the  orthogonal projection matrix  on the kernel
 of the $i$th block row in $M$. Under  exactly the same the conditions as those stated in Theorem \ref{mainer}, the $x_i$ so obtained will
 all converge exponentially fast to the desired least squares solution.

\subsection{Generalization}

 The idea just illustrated by example,  generalizes
in a straight forward way  to any $m$ agent network. The first step  would be to pick any  $m$ vertex graph tree graph
$\mathbb{T}$  and orient it. Agent $i$'s {\em augmented state} would then be    of the form
 $\bar{x}_i = \matt{x_i' & x_{i1}' &x_{i2}' &\ldots &x_{i(m-1)}'}'$ where all $x_{ij}\in\R^n$.
 Instead of solving  $A_ix_i = b_i$,  agent $i$ would be tasked with solving
$\matt{A_i'A_i &h_i\otimes I}\bar{x}_i = A_i'b_i$ where $h_i$ is the $i$th row of the $m\times (m-1)$  incidence matrix of $\mathbb{T}$. At the same time, all $m$ agents would be expected to reach
a consensus in which all $\bar{x}_i$ are equal. Were a consensus reached at a value $\bar{x}=\matt{x' &y_1' &y_2'& \ldots &y_m'}'$, then $\bar{x}$ would have to satisfy the equation $M\bar{x} = q$ where
$$M = \matt{A_1'A_1 &  \cr    \vdots  & H\otimes I \cr A_m'A_m &}\;\;\;{\rm and}\;\;\; q = \matt{A_1'b_1 \cr \vdots \cr A_m'b_m}.$$
We claim that a solution to $M\bar{x} = q$ must exist and that the  sub-vector $x$  within $\bar{x}$  is the solution to the least squares problem. To understand why, first note
 that the block rows of $H\otimes I$ sum to zero because the rows of $H$ sum to zero.
 Thus if $E$ is product of elementary row matrices which
adds the first $(m-1)$ block rows of $H\otimes I$ to the last, then
$E(H\otimes I)$ must be of the form
$$E(H\otimes I) = \matt{D\cr 0}_{nm\times (m-1)n}$$
where $D$ is a square matrix.  Moreover $D$ must be nonsingular because the $\rank E(H\otimes I) = \rank H\otimes I$
and $\rank H\otimes I = (m-1)n$.  This last rank identity is a consequence of the fact that the rank of an incidence matrix of an $m$ vertex  connected graph, namely  $\rank H$,  equals $m-1$.

Next observe that the set of solutions  to $M\bar{x} = q$ is the same as the set of solutions to $EM\bar{x} = Eq$.
But
$$EM = \matt{A_1'A_1 &  \cr    \vdots  & D \cr A_{m-1}'A_{m-1} &\cr A'A & 0}\;\;\;{\rm and}\;\;\; Eq = \matt{A_1'b_1 \cr \vdots \cr A_{m-1}'b_{m-1}\cr A'b}.$$ Moreover  $EM$ is obviously nonsingular so a solution  to $EM\bar{x} = Eq$ and consequently $M\bar{x} = q$ exists.
 Note in addition, that since such a solution must also satisfy
  $EM\bar{x} = Eq$, $x$ must satisfy $A'Ax = A'b$.
  Therefore $x$ solves the least squares problem.

 We have just shown that  if  each agent $i$ updates its augmented state $\bar{x}_i(t)$  along a path for which
 $\matt{A_i'A_i &h_i\otimes I}\bar{x}_i(t) = A_i'b_i$,  so that  $\bar{x}_i(t)$  reaches a limit which agrees with the augmented states of all other agents, then the limiting value of  the sub-vector $x_i(t)$ will solve the least squares problem. The agent update equations for accomplishing this   are  identical to those in \rep{a1} except that in place of  and $x_i$ and $P_i$, agent $i$ would use the
 $\bar{x}_i$ and  $\bar{P}_i$  where $\bar{P}_i$  is the  orthogonal projection matrix  on the kernel
 of $\matt{A_i'A_i & h_i\otimes I}$.  Under  exactly the same the conditions as stated in  Theorem \ref{mainer}, the $x_i$ so obtained will
 all converge exponentially fast to the desired least squares solution.

Although the algorithm just described   solves the  distributed least squares problem, it has several shortcomings. First,
there must be a network wide design step in which $\mathbb{T}$ is specified; this conceivably can be accomplished in a distributed manner. 
Second,  the size of the augmented state vector of each agent  is $nm$ which does not scale well with the number of agents in the network.
 It is possible to significantly improve on the scaling problem if  neighbor relations are time invariant and there is bi-directional communication  between neighbors. How to do this will be addressed in another paper.

\section{Concluding Remarks} In this paper we have described a distributed algorithm for solving a solvable linear equation and given necessary and sufficient conditions for it to generate a sequence of estimates which converge to a solution exponentially fast. For the case when the equation admits a unique solution, we have derived an
expression for a worst case geometric convergence rate.  We have shown that with minor modification, the algorithm
can track the solution to $Ax=b$  if $A$ and $b$ change with time, provided the rates of change of these two matrices are sufficiently  small.  We have show that the same algorithm can function asynchronously provided there are no communications delays and we have sketched a new idea for  obtaining least squares solutions to $Ax=b$ which
can be used even if $Ax=b$ has no solution.

We have left a number of issues opened for future research.  One is to figure out what the relationship is between the parameter $\rho$ which appears in the convergence rate bound $\rho$, and a conditioning number of $A$. Another
is to more tightly quantify the relationship between the variations in $A$ and $b$ in the event they are time varying, and the tracking error $e$.  Yet another is to modify the least squares algorithm discussed in \S\ref{LS}
to reduce the amount of information which needs to be transmitted between agents. This last issue will be addressed in a future paper.

\bibliographystyle{unsrt}
\bibliography{my,steve}

%
%

\end{document}